\documentclass[preprint,sort&compress,12pt]{elsarticle}

\usepackage[top=1in, bottom=1in, left=1in, right=1in]{geometry}

\usepackage{graphicx}
\usepackage{amsmath}

\usepackage{natbib}
\usepackage{amssymb}

\usepackage{amsthm}

\usepackage{lineno}
\usepackage{subfig}
\usepackage{algorithm}
\usepackage{algpseudocode}
\usepackage{xcolor}
\usepackage[colorinlistoftodos]{todonotes}

\biboptions{sort,compress} 

\usepackage{xcolor}

\newcommand*\patchAmsMathEnvironmentForLineno[1]{%
  \expandafter\let\csname old#1\expandafter\endcsname\csname #1\endcsname
  \expandafter\let\csname oldend#1\expandafter\endcsname\csname end#1\endcsname
  \renewenvironment{#1}%
     {\linenomath\csname old#1\endcsname}%
     {\csname oldend#1\endcsname\endlinenomath}}%
\newcommand*\patchBothAmsMathEnvironmentsForLineno[1]{%
  \patchAmsMathEnvironmentForLineno{#1}%
  \patchAmsMathEnvironmentForLineno{#1*}}%
\AtBeginDocument{%
\patchBothAmsMathEnvironmentsForLineno{equation}%
\patchBothAmsMathEnvironmentsForLineno{align}%
\patchBothAmsMathEnvironmentsForLineno{flalign}%
\patchBothAmsMathEnvironmentsForLineno{alignat}%
\patchBothAmsMathEnvironmentsForLineno{gather}%
\patchBothAmsMathEnvironmentsForLineno{multline}%
}

\usepackage{graphicx}
\usepackage{amssymb}
\usepackage{amsthm}
\usepackage{bbm}
\usepackage{bm}
\usepackage{lineno}
\usepackage{url}
\usepackage{listings}
\usepackage[colorlinks=true]{hyperref}

\usepackage{color,soul}
\usepackage{mathtools}

\definecolor{lightblue}{rgb}{.90,.95,1}
\definecolor{darkgreen}{rgb}{0,.5,0.5}

\definecolor{lightgreen}{rgb}{.90,1,0.90}

\newcommand{\bstau}{\boldsymbol{\tau}}

\newcommand{\bs}[1]{\boldsymbol{#1}}

\usepackage{gensymb}
\usepackage{array}
\usepackage{changes}
\usepackage{multirow}
\usepackage{enumerate}
\newcolumntype{P}[1]{>{\centering\arraybackslash}m{#1}}

\newcolumntype{L}[1]{>{\raggedright\let\newline\\\arraybackslash\hspace{0pt}}m{#1}}
\newcolumntype{C}[1]{>{\centering\let\newline\\\arraybackslash\hspace{0pt}}m{#1}}
\newcolumntype{R}[1]{>{\raggedleft\let\newline\\\arraybackslash\hspace{0pt}}m{#1}}

\usepackage{changes}
\definechangesauthor[name={Reviewer 1}, color = blue]{R1}
\definechangesauthor[name={Reviewer 2}, color = brown]{R2}
\definechangesauthor[name={All reviewers}, color = purple]{All}
\definechangesauthor[name={Editor}, color = red]{Editor}
\definechangesauthor[name={Jin-Long Wu}, color = olive]{Author}
\definechangesauthor[name={Heng Xiao}, color = red]{hx}

\DeclareMathOperator*{\argmin}{arg\,min}

\graphicspath{ {./figs/} }

\journal{Physical Review Fluids}

\begin{document}

\begin{frontmatter}

\title{Physics-Informed Machine Learning Approach for Augmenting Turbulence Models: A Comprehensive Framework}

\tnotetext[disclosure]{Disclosure: Parts of this manuscript have been presented in the Proceedings of the CTR Summer Program 2016~\cite{wang16towards} and in an arXiv preprint  arXiv:1701.07102. However, no parts of this paper have been officially published in archived journals.}

\author{Jin-Long Wu}
\ead{jinlong@vt.edu}
\author{Heng Xiao\corref{corr}}
\ead{hengxiao@vt.edu}
\author{Eric Paterson\corref{egp}}
\address{Department of Aerospace and Ocean Engineering, Virginia Tech, Blacksburg, VA 24060, USA}
\cortext[corr]{Corresponding author. +1 540 235 6760}

\begin{abstract}
Reynolds-averaged Navier--Stokes (RANS) equations are widely used in engineering turbulent flow simulations. However, RANS predictions may have large discrepancies due to the uncertainties in modeled Reynolds stresses. Recently, Wang et al. demonstrated that machine learning can be used to improve the RANS modeled Reynolds stresses by leveraging data from high fidelity simulations (Physics informed machine learning approach for reconstructing Reynolds stress modeling discrepancies based on DNS data. Physical Review Fluids. 2, 034603, 2017). However, solving for mean flows from the improved Reynolds stresses still poses significant challenges due to potential ill-conditioning of RANS equations with Reynolds stress closures. Enabling improved predictions of mean velocities are of profound practical importance, because often the velocity and its derived quantities (QoIs, e.g., drag, lift, surface friction), and not the Reynolds stress itself, are of ultimate interest in RANS simulations.
To this end, we present a comprehensive framework for augmenting turbulence models with physics-informed machine learning, illustrating a complete workflow from identification of input features to final prediction of mean velocities. This work has two innovations. First, we demonstrate a systematic procedure to generate mean flow features based on the integrity basis for mean flow tensors. Second, we propose using machine learning to predict linear and nonlinear parts of the Reynolds stress tensor separately. Inspired by the finite polynomial representation of tensors in classical turbulence modeling, such a decomposition is instrumental in overcoming the ill-conditioning of RANS equations. Numerical tests demonstrated merits of the proposed framework. 
\end{abstract}

\end{frontmatter}


\section{Introduction}
\label{sec:intro}

Numerical simulations based on Reynolds-averaged Navier--Stokes (RANS) models are still the work-horse tool in engineering design involving turbulent flows. However, predictions from RANS simulations are known to have large discrepancies in many flows of engineering relevance, including those with swirl, pressure gradients, or mean streamline curvature~\cite{craft96development}. It is a consensus that the dominant cause for such discrepancies is the RANS-modeled Reynolds stresses~\cite{oliver09uncertainty}. In light of the long stagnation in traditional turbulence modeling, researchers~\cite{tracey13application, tracey15machine,ling16machine,ling16reynolds,wang17physics-informed} explored machine learning as an alternative to improve RANS modeling by leveraging data from high-fidelity simulations.

\subsection{Data-driven methods for reducing model discrepancies in RANS simulations}

Data-driven methods have been devised to calibrate the model form uncertainties in RANS simulations based on optimization~\cite{dow11quanti} and Bayesian inference approaches~\cite{dow11quanti,parish16paradigm,singh16using,xiao15quantifying,wu16bayesian}. However, these data-driven calibration approaches inferred the model discrepancies of a given flow and thus lack the generalization capabilities for predicting flows with vastly different geometries from the calibration flow. Therefore, follow-on works of these researchers build data-driven turbulence models in the mean flow features space (as opposed to physical space). Such an approach enables prediction for flows in different geometries yet with similar physics (e.g., curved pipes and wing--body juncture, both featuring secondary flows driven by Reynolds stress anisotropies). Duraisamy and 
co-workers~\cite{tracey13application,duraisamy15new,singh17machine} used non-dimensional flow variables as the input features and a multiplicative correction term in the Spalart--Allmaras model as the machine learning output. Their machine-learning-augmented model demonstrated good generalization capabilities within a class of flows around airfoils~\cite{singh17machine}. Ling et al.~\cite{ling16machine} pointed 
out the importance of embedding the tensorial invariance properties in the machine learning process and used this approach to predict the Reynolds stress with a deep neural network~\cite{ling16reynolds}. Wang et al.~\cite{wang17physics-informed} build a machine-learning model to predict the discrepancies in the RANS modeled Reynolds stresses. Encouraging results have been demonstrated in prediction of Reynolds stresses in two sets of canonical flows (separated flows over periodic hills and secondary flows in a square duct). However, the machine-learning-predicted Reynolds stress leads to large error of solved mean velocities when substituted into the RANS equations. Such an ill-conditioning issue is a common challenge for data-driven Reynolds stress models that must be addressed to unleash the power of such models.

A distinctly different approach of data-driven modeling is pursued by Weatheritt and Sandsberg~\cite{weatheritt16novel,weatheritt17development}, who used symbolic regression and gene expression programming to develop algebraic Reynolds stress models. To some extent, their approach is a combination of traditional modeling and data-driven modeling methods reviewed above. Specifically, while data-driven methods are used to obtain their model, the end product is an algebraic Reynolds model in the traditional sense. As such, the ill-conditioning issue for their model would be similar to the traditional models with explicit analytical forms and not the data-driven models.

\subsection{Conditioning of data-driven Reynolds stress models}
\label{sec:intro-condition-rans}

In this work we refer to solving the RANS equations for mean velocities with a given Reynold stress field  as ``propagation'', which is a critical component in data-driven turbulence modeling. Poroseva et al.~\cite{poroseva16on} referred to such simulation as ``RANS--DNS simulations''. Admittedly, both terminology could cause potential confusions, which thus warrants the explicitly clarification here.

Recently, several researchers have observed that small errors in the Reynolds stresses can be amplified to large errors in the mean velocities when solving the RANS equations with specified Reynolds stresses.
Thompson et al.~\cite{thompson16methodology} propagated Reynolds stresses in  channel flows at a wide range of frictional Reynolds numbers ($Re_\tau=180$ to 5200) to mean velocities by using several reputed DNS databases.
They reported that the propagated mean velocities can deviate significantly from the mean velocities from the DNS, especially for flows at high Reynolds numbers (notably $Re_\tau = 5200)$.  Poroseva et al.~\cite{poroseva16on} also made similar observations, and Poroseva~\cite{poroseva2018personal} further pointed out that the discrepancies between the propagated velocities and DNS velocities were observed for flows at Reynolds numbers as low as $Re_\tau=395$, depending on the dataset used. Considering that errors in DNS Reynolds stress are typically less than $0.5\%$~\cite{lee15direct,wu17on}, these exercises of propagating DNS Reynolds stresses to mean velocities thus represent an ideal scenario for data-driven turbulence models with negligible modeling errors. Wu et al.~\cite{wu17on} explained such observations  by pointing out that RANS equations with Reynolds stress closure models can be ill-conditioned. They further proposed a condition number function defined based on the local velocities to quantify the ill-conditioning.  For plane channel flows, the local condition number does increase with Reynolds number, which thus explained the  increased ill-conditioning with increasing Reynolds number. In contrast, the traditional, matrix-based conditional number was not able to explain such observations. A physical explanation is that viscous stresses are negligible at high Reynolds numbers and the mean velocity is determined by the dependence of the Reynolds stresses on the mean velocity gradients. Therefore, obtaining the mean velocity would fail from an \textit{a priori} specification of the Reynolds stresses.
A unique issue associated with data-driven modeling is that it can be difficult or even impossible to treat the Reynolds stress implicitly as in traditional models, and in such cases segregated solvers are the only option. More detailed discussion can be found in~\cite{wu17on}.

Similar ill-conditioning issue also exists in traditional Reynolds stress models (RSM), where Reynolds stresses are obtained by solving transport equations.  To enhance the stabilities, Jarklic and co-workers~\cite{basara03new,maduta17improved} blended $\bm{\tau}_{\textrm{RSM}}$ obtained from solving the Reynolds stress transport equation with that given by a linear eddy viscosity model (LEVM), i.e., $\bm{\tau}=\alpha \bm{\tau}_{\textrm{RSM}}+(1-\alpha) \bm{\tau}_{\textrm{LEM}}$.  However, the specification of a blending factor $\alpha$ is largely \textit{ad hoc} and lacks physical basis. In this work, we aim at introducing a more rigorous, physics-based implicit treatment in the context of data-driven Reynolds stress models.

\subsection{Data-driven closure modeling beyond RANS simulations}
In addition to RANS modeling as reviewed above, data and machine learning have been used to provide closures for (1) the subgrid-scale (SGS) fluxes for in LES, (2) the inter-phase momentum fluxes in multiphase flow simulations~\cite{ma15using,ma16using}, and (3) the unresolved boundary layer physics in potential flow simulations~\cite{marques17data}. Among these, researchers reported ill-conditioning issues in data-driven SGS models in LES that are similar to the ill-conditioning issue in the context of RANS modeling discussed above. For example, Gamahara et al.~\cite{gamahara17searching} used neural network to model the subgrid-scale stress in a turbulent channel flow. Compared to the predictions of Smagorinsky models, the machine learning model predicted better SGS stresses but the less satisfactory mean velocities. This observation clearly highlights the gap between \textit{a priori} and \textit{a posterior} performances in assessment of turbulence models, particularly in the context of data-driven turbulence modeling. Furthermore, Durieux~\cite{durieux15exploring} reported that the LES with neural network predicted SGS models become unstable if velocity-derived variables are chosen as neural network inputs, suggesting possible error amplification in the propagation of  SGS stresses to mean velocities.

Note that the ill-conditioning issue only emerges if the data-driven SGS model is explicitly substituted into filtered transport equations to solve for velocities or other quantities of interest (QoIs). Many other works that focus on the data-driven prediction of SGS terms (i.e., \textit{a priori} tests) did not encounter this problem.  For example, Vollant et al.~\cite{vollant14optimal,vollant17subgrid} modeled SGS scalar flux by using neural networks based on optimal estimation. King et al.~\cite{king16autonomic} proposed a fully adaptive, self-optimizing SGS closure and demonstrated superior \textit{a priori} performance than traditional dynamic SGS models. Maulik and San~\cite{maulik17neural} trained a neural network to represent the deconvolution of flow quantities from filtered flow field. They also demonstrated excellent \textit{a priori} performance in several canonical, boundary-free flows.  All these promising works have the potential of becoming data-driven SGS closures for stresses or scalar fluxes in LES.  However, similar ill-conditioning issue in the context of RANS modeling as outlined in Section~\ref{sec:intro-condition-rans} still needs to be addressed if the successes in these \textit{a priori} tests are to be translated into \textit{a posteriori} tests.

\subsection{Summary and novelty of present contribution}
\label{sec:novelty}

In the present work, we demonstrate a \emph{systematic} approach in choosing the input feature variables for machine learning in the context of turbulence modeling. Specifically, we first identify a set of vectorial or tensorial mean flow variables, e.g., strain-rate tensor, rotation-rate tensor, pressure gradient, and turbulence kinetic energy gradient. Choosing these quantities as inputs for machine learning has clear physical justifications and are supported by the practice in traditional turbulence modeling. Subsequently, we construct an invariant basis set from these variables based on the tensor representation theorem. While this approach was first proposed by Ling et al.~\cite{ling16machine} and is not new in the present work, the application in turbulence modeling is challenging and could serve as good illustration for researchers in many other fields where the physical quantities are described by a large number of vectors and tensors. This procedure is a clear improvement compared to earlier works with \textit{ad hoc} choice of many scalar variables~\cite{ling15evaluation,wang17physics-informed}.

Moreover, we propose a data-driven, machine learning based turbulence modeling framework where the Reynolds stresses are decomposed into linear and nonlinear parts and then learned separately form DNS data. This decomposition allows implicit treatment of linear term of the Reynolds stress, which enhances the model conditioning in solving the RANS equations for mean velocity field without \textit{ad hoc} blending as used in traditional turbulence modeling~\cite{basara03new,maduta17improved}. More importantly, such a decomposition clearly reflects the respective roles of linear and nonlinear terms in turbulent models. Specifically, the linear term is by far the dominant term in almost all turbulent flows in engineering practice, which partly explains the wide spread use of linear eddy viscosity models in engineering CFD. On the other hand, the nonlinear terms can emerge as important factors in specific flows (e.g., swirling, jet impingement, and juncture flows~\cite{craft96development,apsley01investigation}). Hence, treating them separately helps the machine learning algorithms distinguish these two terms.  With numerical examples, we show that such a treatment enabled accurate prediction of mean velocities with a data-driven Reynolds stress model.

The rest of this paper is organized as follows. Section~\ref{sec:PIML} summarizes the machine-learning-assisted turbulence modeling framework of Wang et al.~\cite{wang17physics-informed} and presents the proposed approach. Section~\ref{sec:results} first highlights the ill-conditioning issues of data-driven Reynolds stress models and further demonstrates the merits of the proposed machine learning framework in a posteriori tests of different kinds of flows. Section~\ref{sec:discussion} discusses the potentials and limitations of the data-driven turbulence models. Finally, Section~\ref{sec:conclusion} concludes the paper.

\section{Methodology}
\label{sec:PIML}
Taking incompressible turbulent flows as an example, the RANS momentum equations are:
\begin{equation}
  \frac{\partial \mathbf{U}}{\partial t}
  + \mathbf{U}  \cdot \nabla 
   \mathbf{U}
  +  \nabla p -
  \nu \nabla^2  \mathbf{U} =  \nabla \cdot \boldsymbol{\tau}
 \label{eq:rans-momentum}
\end{equation}
where $\mathbf{U}$, $p$, $\nu$ are the mean velocity, mean pressure (normalized by density), and viscosity, respectively.  The Reynolds stress $\boldsymbol{\tau}$ accounts for the momentum flux due to unresolved turbulence and needs closure modeling. A turbulence model aims to close the RANS equations by constructing a mapping from the mean velocity field $\mathbf{U}$ to the Reynolds stress field $\boldsymbol{\tau}$. 

In view of the inaccuracy in RANS modeled Reynolds stresses as a critical bottleneck in the prediction accuracy of CFD simulations, Wang et al.~\cite{wang17physics-informed} proposed a machine learning model for predicting Reynolds stress discrepancies by training on DNS data from similar flows. Specifically, a functional mapping $\mathbf{q} \mapsto \Delta \bm{\tau}$ from mean flow features $\mathbf{q}$ (obtained from RANS simulations) to Reynolds stress discrepancies $\Delta \bm{\tau}$ is built by using machine learning, with the discrepancy defined as the difference between RANS predicted and DNS Reynolds stresses, i.e., $\Delta \bm{\tau} \equiv \bm{\tau}^{\textrm{DNS}}-\bm{\tau}^{\textrm{RANS}}$. Although an improved prediction of Reynolds stresses was achieved, they reported that the mean flow velocity is sensitive to the errors in Reynolds stress prediction. Since it is usually the mean velocity field and the derived quantities of interest (e.g., drag and lift) required in engineering applications, it is important to understand the error amplification in solving for the mean velocity field. The main reason is that substituting the modeled Reynolds stress explicitly into RANS equations may lead to ill-conditioned RANS equations, especially in high Reynolds number flows~\cite{thompson16methodology,wu17on}. Moreover, the choice of mean flow features lacks a systematic procedure, which raises questions on the completeness and redundancy on the set of flow features included in the machine learning. The present work aims to address these challenges.

\subsection{Overview of the machine-learning scheme}
\label{sec:framework-overview}
It has been recognized that all algebraic Reynolds stress and eddy viscosity models can be written in the following general form~\cite{pope75more}:
\begin{equation}
\label{eq:bij-pope}
\begin{aligned}
\mathbf{b} (\mathbf{S}, \bm{\Omega}) 
= &\sum_{n=1}^{10}G^{(n)}\mathbf{\mathcal{T}}^{(n)} \\
= &  G^{(1)} \mathbf{S} + G^{(2)} (\mathbf{S} \bm{\Omega} - \bm{\Omega} \mathbf{S}) + G^{(3)} (\mathbf{S}^2-\frac{1}{3}\textrm{tr}(\mathbf{S}^2)\mathbf{I})  \\
&  + G^{(4)} (\bm{\Omega}^2-\frac{1}{3}\textrm{tr}(\bm{\Omega}^2)\mathbf{I}) + \cdots \text{(high order terms)}
\end{aligned}
\end{equation}
where $\textrm{tr}(\cdot)$ denotes the trace, $\mathbf{I}$ denotes the identity matrix, $\mathbf{b}$ is the deviatoric part of Reynolds stress tensor, $\{\mathbf{\mathcal{T}}^{(n)}\}_{n=1}^{10}$ is the tensorial basis formed from strain-rate tensor $\mathbf{S} = \frac{1}{2}\left(\nabla \mathbf{U} + (\nabla \mathbf{U})^T\right)$ and rotation tensor $\mathbf{\Omega} = \frac{1}{2}\left(\nabla \mathbf{U} - (\nabla \mathbf{U})^T\right)$. In particular, $\mathbf{\mathcal{T}}^{(1)}=\mathbf{S}$ and thus $G^{(1)}\mathbf{\mathcal{T}}^{(1)}$ represents the linear part (with respective to $\mathbf{S}$) of the anisotropy tensor $\mathbf{b}$.

Inspired by this general form of algebraic Reynolds stress models, we separate the anisotropy stress tensor $\bm{b}$ into linear and nonlinear parts:
\begin{equation}
\label{eq:b}
\mathbf{b}=\nu_t^{L} \mathbf{S} +\mathbf{b}^\perp
\end{equation}
where linear part $\nu_t^{L} \mathbf{S}$ (co-axial with $\mathbf{S}$ corresponds to the term $G^{(1)} \mathbf{S}$ in Eq.~\eqref{eq:bij-pope}), and $\mathbf{b}^\perp$ represents the sum of the non-linear terms. It is similar to the tensorial expansion of anisotropy stress tensor $\mathbf{b}$ in Eq.~\eqref{eq:bij-pope}, while all the nonlinear terms are lumped into $\mathbf{b}^\perp$. The linear term can be treated implicitly to enhance the conditioning when solving the RANS equations. More details of numerical procedure for solving the RANS equations are detailed in~\ref{sec:workflow}. As discussed above in Sec.~\ref{sec:novelty}, this decomposition is more than just a numerical implicit treatment but has clear physical justifications. 

In order to compute the two terms in Eq.~\eqref{eq:b} from a given Reynolds stress and strain tensor, we introduce an optimal eddy viscosity that minimizes the discrepancy between the anisotropy Reynolds stress tensor and its linear part, i.e., 
\begin{equation}
\nu_t^L=\argmin_{\nu_t}\|\mathbf{b}-\nu_t\mathbf{S}\|
\end{equation}
where $\|\cdot\|$ denotes the Frobenius norm of a matrix, e.g., $\|\mathbf{S}\|=\sqrt{S_{ij}S_{ij}}$. Based on this definition, the optimal eddy viscosity $\nu_t^L$ can be computed by projecting the anisotropy stress tensor on the strain rate tensor:
\begin{equation}
\label{eq:nut}
\nu_t^{L}=2\frac{\mathbf{b}:\mathbf{S}}{\|\mathbf{S}\|\|\mathbf{S}\|}
\end{equation}
where $\mathbf{b}:\mathbf{S}=b_{ij}S_{ij}$ denotes tensor double dot product.

The nonlinear term $\mathbf{b}^\perp$ in Eq.~\eqref{eq:b} could be important even for simple shear flows in the near wall region, since the linear term incorrectly predicts isotropic normal stresses. For more complex flows, e.g., swirling and impinging, neglecting this nonlinear term can cause the model to miss important flow physics completely. In this work, we use machine learning techniques and existing DNS database to build regression functions that predict the optimal eddy viscosity $\nu_t^L$ and the nonlinear part $\mathbf{b}^\perp$ of the anisotropy Reynolds stress tensor. In machine learning terminology the flows used to build regression functions are referred to as the \textit{training flows}, and the flow to be predicted is referred to as the \textit{test flow}. The detailed workflow of building these regression functions via machine learning is presented in \ref{sec:workflow}.

\subsection{Construction of mean flow features as inputs of machine learning}
\label{sec:meth-input}
The construction of input features is among the most critical considerations when using machine learning for physical problems. First, the choice of input and output variables must be  physically motivated and justified to ensure that the function learned from the data has physical meaning. Second, the variables must be normalized properly to ensure extrapolative capabilities of the learned function. Finally, the learned function should ideally be objective with function form invariances under transformations of the coordinate system and the reference frame.  Our perspective is that almost all principles that are observed in traditional turbulence  modeling~\cite[see, e.g.,][]{spalart15philosophies} should be equally respected in data-driven turbulence modeling. These three considerations in the present framework are presented below. 
 
 \subsubsection{Physical consideration in the choice of mean flow feature variables}
 \label{sec:meth-input-physics}
 
The general form of nonlinear turbulent-viscosity model in Eq.~\ref{eq:bij-pope} assumes a universal functional mapping from the strain-rate tensor $\mathbf{S}$ and the rotation-rate tensor $\mathbf{\Omega}$ to the Reynolds stress $\bm{\tau}$:
\begin{equation}
\label{eq:pope-nut}
\bm{\tau}=\bm{\tau}(\mathbf{S},\mathbf{\Omega})
\end{equation}
We note that there are at least two aspects of missing physics in this assumption. First, the turbulence is also influenced by pressure gradient. For example, turbulence would be suppressed under strong favorable pressure gradient~\cite{spalart15philosophies}. On the other hand, the general form in Eq.~\ref{eq:pope-nut} assumes equilibrium turbulence, i.e., the turbulence production balances dissipation everywhere in the field. With such an assumption, the Reynolds stress at location $\bm{x}$ only depends on the \emph{local} mean velocity $\mathbf{U}(\bm{x})$, or more precisely, its gradient $\nabla \mathbf{U}(\bm{x})$. However, the convection and diffusion of turbulence exist in many real applications, indicating strong non-equilibrium effects and making this single-point-based turbulent constitutive law invalid~\cite{lumley70toward}. To account for the missing physics outlined above, we also include the pressure gradient $\nabla p$ and the TKE gradient $\nabla k$ in the input, leading to a more general functional mapping from mean flow quantities to the Reynolds stress:
\begin{equation}
\label{eq:raw}
\bm{\tau} =  g(\mathbf{S}, \bs{\Omega}, \nabla p, \nabla k)
\end{equation} 	
where the set of varialbles $\mathcal{Q} = \{\mathbf{S}, \bs{\Omega}, \nabla p, \nabla k \}$ are chosen as input features, which are summarized in Table~\ref{tab:featureRaw}.

\begin{table}[htbp] 
	  \def\arraystretch{1.2}		
	\centering
	\caption{
		Non-dimensional raw mean flow variables used to construct the invariant basis.  
		The normalized feature $\widehat{\alpha}$ is obtained by normalizing the corresponding 
		raw input $\alpha$ with normalization factor $\beta$ according to 
		$\widehat{\alpha} = \alpha / (|\alpha| + |\beta|)$. Notations are as follows: $\mathbf{U}$ is 
		mean velocity vector, $k$ is turbulence kinetic energy (TKE), $\rho$ is fluid density, 
		$\varepsilon$ is the turbulence dissipation rate, $\mathbf{S}$ is the strain-rate tensor, 
		$\bs{\Omega}$ is the rotation-rate tensor, $|\cdot|$ denotes vector norm, $\| \cdot \|$ indicates matrix norm.  }
	\label{tab:featureRaw}
	\begin{tabular}{P{2.5cm} | P{3cm}  P{3.0cm}  P{5.0cm} }	
		\hline
		Normalized raw input $\widehat{\alpha}$  & description & raw input $\alpha$ &
		normalization factor $\beta$  \\ 
		\hline
		$\widehat{\mathbf{S}}$  & strain-rate tensor&
		$\mathbf{S}$ & 
		$\dfrac{\varepsilon}{k}$\\  
		\hline
		$\widehat{\bs{\Omega}}$  & rotation-rate tensor & $\bs{\Omega}$ &
		$\|\mathbf{\Omega}\|$\\ 
		\hline
		$\widehat{\nabla p}$  & pressure gradient &
		$\nabla p$ & $\rho|D\mathbf{U}/Dt|$ \\
		\hline
		$\widehat{\nabla k}$  & TKE gradient & $\nabla k$ & $\dfrac{\varepsilon}{\sqrt{k}}$ \\ 
		\hline					 								
	\end{tabular}
\end{table}

In addition to the tensor set $\mathcal{Q}$, three other features as presented in Table~\ref{tab:feature} are chosen from Ref.~\cite{wang17physics-informed} to further supplement the mean flow features, all of which have clear physical interpretations. First, at the near wall region the viscous effect becomes more important and the local Reynolds number reduces to $O(1)$. Therefore, a low Reynolds number turbulence model is needed for the viscous sublayer in the traditional turbulence modeling. In this work, $q_1$ is an important indicator to inform the wall distance to the machine-learning-assisted turbulence models, leading to a data-driven low-$Re$ model as a counterpart of the traditional low-$Re$ models. Second, features $q_2$ and $q_3$ carry information on the length-scale and time-scale of the turbulence, serving as supplements of the mean flow tensors in the set $\mathcal{Q}$.

\begin{table}[htbp] 	
  \centering
  \caption{
    Supplementary mean flow features used as inputs in the regression.  The normalized feature $q_\beta$ is
    obtained by normalizing the corresponding raw features value $\widehat{q}_\beta$ with normalization
    factor $q^*_\beta$ according to $q_\beta = \widehat{q}_\beta / (|\widehat{q}_\beta| + |q^*_\beta|)$
    except for $\beta = 1$. Notations are as follows: $U_i$ is mean velocity, $k$
    is turbulent kinetic energy (TKE), 
    $\varepsilon$ is the turbulence dissipation rate, $\mathbf{S}$ is the strain rate tensor,
    $d$ is the distance to the
    wall. $\| \cdot \|$ indicates matrix norms.  }
\label{tab:feature}
\begin{tabular}{P{2.0cm} | P{5.0cm}  P{3.0cm}  P{4.0cm} }	
  \hline
  feature ($q_\beta$)  & description & raw feature ($\widehat{q}_\beta$) &
  normalization factor ($q^*_\beta$)  \\ 
  \hline
  $q_1$  & wall-distance based Reynolds number &
  $\min{\left(\dfrac{\sqrt{k}d}{50\nu}, 2\right)}$ & not applicable$^{(a)}$\\
  \hline
  $q_2$  &   turbulence intensity & $k$ &
  $\nu\|\mathbf{S}\|$\\ 
  \hline
  $q_3$  & ratio of turbulent time-scale  to mean strain time-scale &{ $\dfrac{k}{\varepsilon}$ }
  & { $\dfrac{1}{\|\mathbf{S}\|} $ } \\ 
  \hline						 								
\end{tabular}
\flushleft
{\small
  Note: (a)  Normalization is not necessary as the Reynolds number is non-dimensional.}
\end{table}
 
\subsubsection{Normalization of input features}
To ensure non-dimensionality of the raw inputs, the normalization scheme proposed by Ling and Templeton~\cite{ling15evaluation} is adopted. All raw features are normalized by \emph{local} quantities, as is preferred in the practice of traditional turbulence modeling~\cite{spalart15philosophies}. In CFD simulations these would be quantities based on the same grid point as the raw feature variables. The normalization factors for all the raw input variables are listed in Table~\ref{tab:featureRaw}.  Specifically, each element $\alpha$ in the raw input set~$\mathcal{Q}$ is normalized by a corresponding normalization factor $\beta$ based on the following scheme:
\begin{equation}
  \label{eq:normal}
\widehat{\alpha} = \frac{\alpha}{|\alpha| + |\beta|},  
\end{equation}
which ensures that the normalized variable $\widehat{\alpha}$ falls within the range $[-1, 1]$. Note that such a normalization scheme is slightly different from that frequently used in physics and engineering, which would take either the form  $\widehat{\alpha} = \alpha/|\alpha|$ or $\widehat{\alpha} = \alpha/|\beta|$ instead.  The choice in Eq.~\ref{eq:normal} is justified by the practice of machine learning where the inputs are usually normalized to the range $\left[-1,1\right]$ or $\left[0,1\right]$.  This helps avoid clustering of training data along certain directions within the input feature space and  improves the convergence rate in the training process.

\subsubsection{Invariance considerations in the choice of input features}
\label{sec:meth-invariance}
As summarized in Table~\ref{tab:featureRaw}, the raw variables for the mean flow  consist of a finite tensorial set $\mathcal{Q} = \{\mathbf{S}, \bs{\Omega}, \nabla p, \nabla k \}$ with four elements, where $\nabla p$ and $\nabla k$ are transformed to anti-symmetric tensors as detailed in Eq.~\eqref{eq:vector2anti}. As in traditional turbulence modeling, it is equally desirable in data-driven turbulence modeling that the trained functional form $g: (\mathbf{S}, \bs{\Omega}, \nabla p, \nabla k) \mapsto \bm{\tau}$ should be objective.
That is, the function form of $g$ should be invariant under rotational and reflectional transformations of the coordinate system or  Galilean transformation (i.e., translation by a constant velocity) of reference frame.  The function form invariances associated with the rotation and reflection of the coordinate system and the Galilean transformation of the reference frame are referred to as \emph{coordinate rotational invariance}, \emph{coordinate reflectional invariance}, and \emph{Galilean invariance}, respectively.  Our formulation has  rotational invariance and Galilean invariance but not reflectional invariance. However, the lack of reflectional invariance can be remedied by data augmentation, which is a standard  procedure for  pre-processing training data in machine learning.  The three invariance properties of the present formulation and strategies to remedy the lack of reflectional invariance are examined below.

\paragraph{Invariance properties}
If the constructed function relation $\bm{\tau} =  g(\mathbf{S}, \bs{\Omega}, \nabla p, \nabla k)$ is to be valid under arbitrary rotations of the coordinate system, the following relation should be satisfied~\cite{speziale91modelling}:
\begin{equation}
  \label{eq:fturb}
  \mathbf{Q} \boldsymbol{\tau} \mathbf{Q}^T = g (\mathbf{Q}  \mathbf{S} \mathbf{Q}^T,  \mathbf{Q} \boldsymbol{\Omega} \mathbf{Q}^T, \mathbf{Q}\nabla p, \mathbf{Q} \nabla k) 
\end{equation}
for any rotation matrix $\mathbf{Q}$, where $\mathbf{Q}$ is an orthogonal matrix (i.e., $\mathbf{Q}^{T} = \mathbf{Q}^{-1}$) with determinant equaling to 1. 
The rotational invariance of the learned function $g$ as stated in Eq.~\eqref{eq:fturb} can be guaranteed by choosing invariant inputs and outputs in the learning process, specifically, by choosing the \emph{minimal integrity bases} for the set $\{\mathbf{S}, \bs{\Omega}, \nabla p, \nabla k\}$ and the invariants of $\bm{\tau}$ as inputs and outputs, respectively. A minimal integrity bases is the minimal set of invariants that can represent all the polynomial invariants associated with a tensorial set under the designated transformation (rotation here). The \emph{Hilbert basis theorem} states that a minimal integrity basis for a finite tensorial set has finite number of invariants~\cite{spencer62isotropic}. Specifically for the set $\mathcal{Q} = \{\mathbf{S}, \bs{\Omega}, \nabla p, \nabla k\}$ of second-order tensors considered here\footnote{Vectors such as $\nabla p$ and $\nabla k$ can be first transformed to the corresponding anti-symmetric tensors based on Eq.~\eqref{eq:vector2anti} in~\ref{sec:list-of-basis}}, 
the minimal integrity basis consists of all the traces of the independent matrix products that can be formed from the tensors according to Cayley-Hamilton theorem~\cite{johnson16handbook}, which amount to 47 invariants (see Table~\ref{tab:basis} for details). Note that choosing the invariant tensorial bases, rather than the raw tensorial variables, as inputs and output of the machine learning only guarantees the rotational invariance of the learned function. Galilean invariance needs to be independently achieved by ensuring each raw variable and their normalization factors  to be Galilean invariant, which is discussed below.

Galilean invariance states that the laws of motion are the same in all frames with constant velocities, which is an important prerequisite of any turbulence model~\cite{spalart15philosophies} and should be equally satisfied by traditional or data-driven models.  Therefore, the velocity \textit{per se} is usually not a valid term in a model as mentioned in~\cite{spalart15philosophies}, because the velocity is not Galilean invariant.  In contrast, the velocity \emph{gradient} $\nabla \mathbf{U}$ and thus its symmetric and anti-symmetric parts ($\mathbf{S}$ and $\boldsymbol{\Omega}$, respectively) are all Galilean invariant and thus are valid terms to be included in a turbulence model. Similarly, the gradient of pressure $\nabla p$ and the gradient of kinetic energy $\nabla k$ are both Galilean invariant (and does not depends on the choice of reference pressure).  Finally, it is straightforward to show that a term involving only Galilean invariant quantities is also Galilean invariant. Based on the general principles outlined above, the machine learning inputs and outputs in the present formulation are all Galilean invariant.  Specifically, the raw inputs in Table~\ref{tab:featureRaw} are all Galilean invariants since they only involves spatial gradients, e.g., $\mathbf{S}$ and $\mathbf{\Omega}$. The raw inputs in Table~\ref{tab:feature} are all Galilean invariant since they only involves scalar quantities that are Galilean invariant.  The normalization factors in both tables are all Galilean invariant as well. In particular, it can be shown that the normalization factor $\rho |D \mathbf{U}/Dt|$ is Galilean invariant, and the details are presented in \ref{app:gali-inv}.

Finally, the invariants associated with the anti-symmetric tensors are only rotational invariants but not reflection invariants. The main motivation for using these anti-symmetric tensors is that the machine learning outputs include the quantification of 3-D rotation of eigenvectors of stress tensors that does not have reflection invariance. The details are discussed as follow:

\begin{itemize}
\item \textit{Motivation of the chosen outputs:} Our framework aims to augment traditional turbulence models, instead of completely replacing them, by using machine learning to predict the discrepancies between RANS modeled and true Reynolds stresses. Such discrepancies are parameterized by the scaling of the Reynolds stress tensor along its eigenvectors and the rotation of the eigenvectors. Although the scaling factors are all scalars and invariants under coordinates system transformation, the rotation of the eigenvectors does not have reflection invariance. For instance, the sign of the angle between any two vectors depends on the defined direction of the rotation axis.

\item \textit{Reasons of including inputs without reflection invariance:} If we only introduce machine learning inputs with both rotational and reflection invariance, the similar inputs would correspond to totally different outputs of tensor rotations (potentially with flipped signs) in the training dataset. Therefore, the functional form $\mathcal{F}: \bm{q} \mapsto \Delta \bm{\tau}$ would be a noisy function with spikes due to the sign flipping learned from the training data. Such a machine learning model would be nonphysical with diminishing predictive capabilities.

\item \textit{Consistency between outputs and inputs:} We adopted the invariants of pseudo-tensors constructed from the TKE gradient and the pressure gradient as a part of machine learning inputs. Similar to the invariants corresponds to a 3-D rotation, the invariants of these pseudo-tensors are also rotational invariants but not reflection invariants. Therefore, the sign flipping of the machine learning outputs can be distinguished by different corresponding inputs and the unphysical noisy behavior of the functional form $\mathcal{F}$ would not exist. The main purpose here is to ensure a consistent framework to predict tensor rotations, which is critical in augmenting the traditional tensor models.
\end{itemize}

Indeed, it would be more elegant to include only objective inputs and output in the machine learning, as the coordinate transformation invariance is a basic requirement in turbulence modeling. Further work is needed in identifying such formulations.It should be noted that the different conventions of coordinate system handedness would lead to different machine learning inputs even with identical training and test flows. An alternative albeit rather inefficient approach is to use training flows under both right-handed and left-handed coordinates system to augment the training data, and it has been demonstrated in~\cite{ling16machine} that the invariance can be learned by such data augmentation.

\paragraph{Data augmentation for achieving invariance in machine learning}
In machine learning the lack of invariance property in the model to be learned from data can be remedied by data augmentation~\cite{ling16machine}.  Specifically, the training dataset is augmented by duplicating them in various transformed coordinate systems before performing the training. That way, the training process would be able to ``see'' the same data in almost all transformed coordinates. Consequently, only the functional forms that are valid (i.e., invariant) in all coordinate systems are learned, and any coordinate- or frame-dependent functional forms would be rejected in the training. However, depending on the invariance to be achieved through data augmentation (as detailed below), this procedure could significantly increase the amount of data and the computational costs for the training and prediction.
\begin{itemize}
  \item In order to achieve reflectional invariance through data augmentation, one only needs to duplicate the data in the reflected coordinate system, which is a moderate two-fold increase in the amount of data.
  \item On the other hand,  achieving three-dimensional rotational invariance  required duplicating the training data in 1000 coordinate systems, as is shown by Ling et al.~\cite{ling16machine}. Therefore, using data augmentation to learn rotational invariance would significantly increase the computational cost and memory consumption, both of which are important considerations in machine learning.
  \item Finally, it is not straightforward to remedy the lack of Galilean invariance with data augmentation, since the translation velocity of the reference frame is unbounded (i.e., it can be any value from $-\infty$ to $\infty$).
  \end{itemize}

In summary, a total of 50 normalized, invariant mean flow field variables (collectively denoted as $\mathbf{q}$) are constructed and used as input features for the machine learning. While such a high-dimensional feature space may appear daunting even for the most experienced experts in turbulence modeling, it is not particularly challenging in the context of modern data science, as many machine learning techniques routinely handle feature spaces of thousands of dimensions or even higher~\cite{bishop06pattern}.

\subsection{Representation of Reynolds stress discrepancy as outputs of machine learning}

Similar to choosing the inputs of machine learning, we represent the Reynolds stress discrepancies with rotationally invariant variables as the outputs of the machine learning. Following~\cite{emory13modeling,wang17physics-informed}, we formulate the Reynolds stress discrepancies 
 as six physically interpretable components (i.e., magnitude, shape, and orientation) based on eigen-decomposition of anisotropic Reynolds stress tensor.
 \begin{equation}
 \label{eq:tau-decomp}
 \boldsymbol{\tau} = 2 k \left( \frac{1}{3} \mathbf{I} +  \mathbf{b} \right)
 = 2 k \left( \frac{1}{3} \mathbf{I} + \mathbf{V} \Lambda \mathbf{V}^T \right)
 \end{equation} 	
 where $k$ is the turbulent kinetic energy, which indicates the magnitude of $\bstau$; $\mathbf{I}$
 is the second order identity tensor; $\mathbf{b}$ is the deviatoric part of $\bstau$; 
 $\mathbf{V} = [\mathbf{v}_1, \mathbf{v}_2, \mathbf{v}_3]$ and $\Lambda = \textrm{diag}[\lambda_1, \lambda_2, \lambda_3]$ with $\lambda_1+\lambda_2+\lambda_3=0$
 are the orthonormal eigenvectors and eigenvalues of $\mathbf{b}$, respectively, indicating its shape
 and orientation. 
 
The eigenvalues $\lambda_1$, $\lambda_2$,
and $\lambda_3$ are transformed to Barycentric coordinates $C_1$, $C_2$, and $C_3$ and then to Cartesian coordinates $\xi$ and $\eta$ as in Refs.~\cite{wang17physics-informed} and~\cite{banerjee07presentation}.

To ensure that the predicted TKE is non-negative, Wang et al.~\cite{wang17physics-informed} introduce the TKE discrepancy $\Delta\log k$ as the logarithm of the ratio of the target TKE ($k^*$) to the 
RANS-simulated TKE ($k^{\textrm{RANS}}$), i.e.,
\begin{equation}
\Delta\log k \equiv \log{\frac{k^*}{k^{\textrm{RANS}}}}.
\end{equation}
This is what we adopted in the current work.
Finally, the unit quaternions are used to represent the transformation from the RANS eigenvectors $\mathbf{V}^{\textrm{RANS}}$ to the target eigenvectors $\mathbf{V}^*$~\cite{wu17representation}:
\begin{equation}
  \label{eq:quaternion}
  \mathbf{h} = \left[\cos{\frac{\theta}{2}}, n_{1}\sin{\frac{\theta}{2}}, n_{2}\sin{\frac{\theta}{2}},   n_{3}\sin{\frac{\theta}{2}}\right]^T
\end{equation}
where $\mathbf{n} \equiv [n_1, n_2, n_3]$ denotes a unique axis of unit vector, and $\theta$ represents the rotation angle such that $\mathbf{V}^*$ can be obtained via rotating $\mathbf{V}$ by $\theta$ about the axis $\mathbf{n}$. This unit quaternions representation is rotational invariant and thus is preferred than the Euler angles representation using by Wang et al.~\cite{wang17physics-informed}.  Note that the current representation of the rotation of Reynolds stress eigenvectors is not reflectional invariant, i.e., the magnitude of each component of the unit quaternion~$\mathbf{h}$ remains the same under the reflection of frame while the sign of each component is not. As mentioned in Section~\ref{sec:meth-input}, the reflection invariance of the trained machine learning model can be achieved by augmenting the training database with reflected coordinates system.
 
In summary, the discrepancies ($\Delta \log k$, $\Delta \xi$, $\Delta \eta$, 
$h_1$, $h_2$, $h_3$), collectively denoted as $\Delta{\bs\tau}$, are used as the machine learning outputs to represent the discrepancies between the target Reynolds stress and the RANS modeled Reynolds stress. Here $h_1$, $h_2$ and $h_3$ denote the first three components of the unit quaternion $\mathbf{h}$. All six variables are invariant under rotations of the coordinate system. In this work, random forest regression is adopted to represent the dependence of these Reynolds stress discrepancies on a large number of scalar inputs as identified in Section~\ref{sec:meth-input}. Therefore, the mean velocity and other quantities (e.g., $k$ and $\varepsilon$) from RANS simulations are used to calculate the machine learning inputs, since it can be expected that the Reynolds stress discrepancies are related to the specific choice of the RANS model. Because of the dependence of the trained machine learning function on the RANS model, we recommend the usage of the same RANS model for both the training flows and the flow to be predicted.

\subsection{Choice of machine learning algorithm and parameters}
In this work, random forest regression~\cite{breiman01random} as implemented in R~\cite{ihaka96r} is used to build functional mappings from the inputs (mean flow features $\mathbf{q}$) to the responses as identified in Sec.~\ref{sec:framework-overview}. Random forest regression is a tree-based ensemble learning method, i.e., the regression outputs are the mean prediction of individual decision trees. In this work, the regression outputs are directly used as explicit values and the detailed formulation of the random forest has no influence upon the convergence of solving mean velocity via RANS equations. An advantage of the random forest regression is that it can provide importance scores for inputs after training, which can be further used to assist the modelers to improve the existing RANS models~\cite{wang17physics-informed}. Random forest has robust performances with only a small set of tuning parameters, which is in contrast to the commonly used neural networks~\cite{liaw02classification}. The number of max features is set as 7, i.e., ($(1+\log_2 n)$, where $n=50$ is the number of input features in this work), based on the recommendation in Ref.~\cite{breiman01random}, ). The number of trees is set as 300. This number is chosen by observing the out-of-bag (OOB) error to avoid possible overfitting on the training sets. We have observed the OOB error with different numbers of trees and this error is not sensitive based on our current setting of the number of trees. An example of using random forest regression to assist RANS modeling is publicly available at \href{url}{https://github.com/xiaoh/turbulence-modeling-PIML}.

The computational costs of machine learning consist of the training cost and the prediction cost. The training cost depends on the amount of the training data, and it should be noted that the training procedure can be done offline and the trained machine learning model can be applied to the prediction of other flows as well. In this work, only one training flow is used at one time and the computational cost of the training procedure is less than the corresponding standard RANS simulation. The computational cost of the prediction procedure is usually negligible compared with the cost of a typical RANS simulation. Therefore, with the trained machine-learning-assisted model, the computational cost for the baseline RANS simulation in combination with the prediction of a given flow is still much lower than  large-eddy simulations (LES) of the same flow.

\section{Numerical results}
\label{sec:results}
Two canonical flows, the flow in a square duct and the flow over periodic hills, are investigated to evaluate the performance of the proposed method. The flow in a square duct is featured by stress-induced secondary flow, and the flow over periodic hills is featured by the massive separation. These two features are challenging for traditional RANS modeling~\cite{huser93direct,breuer09flow}. In this work, we first perform a \textit{propagation} test by using DNS Reynolds stress. The purpose of the \textit{propagation} of DNS Reynolds stress to the mean velocity is to demonstrate the merit of physics-based implicit treatment. In the \textit{a posteriori} test, the Reynolds stress field is modeled by machine learning techniques and propagated to mean velocities to evaluate the predicative capability of the proposed machine-learning-assisted turbulence modeling framework with physics-based implicit treatment.

\subsection{Case setup}
\label{sec:case-setup}
\begin{figure}[htbp]
\centering
\includegraphics[width=0.8\textwidth]{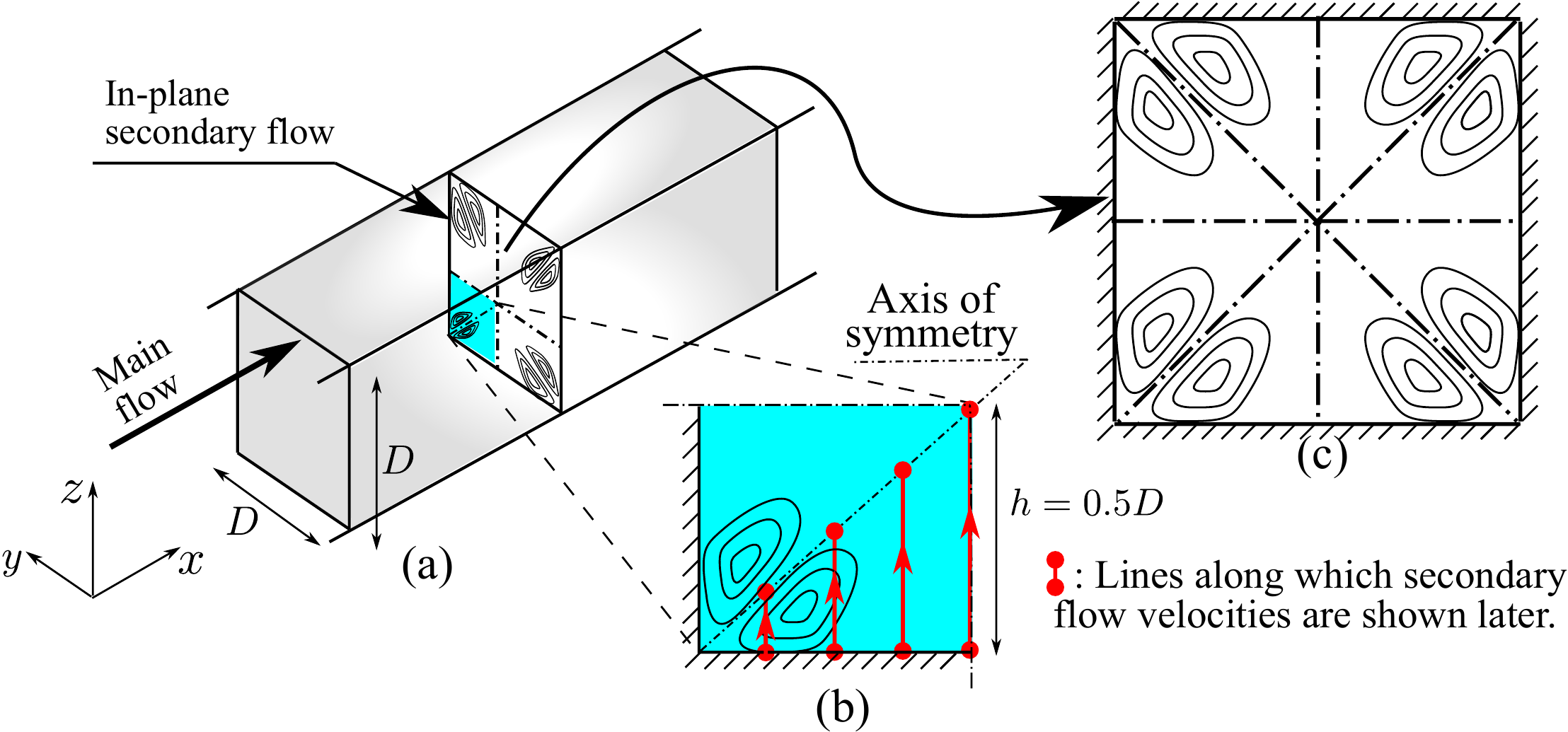}
\caption{Computational domain for the flow in a square duct. The $x$ coordinate represents the streamwise
  direction. Secondary flows induced by Reynolds stress imbalance exist in the $y$--$z$
  plane. Panel (b) shows that the computational domain covers a quarter of the cross-section of the
  physical domain. This is due to the symmetry of the mean flow in both $y$ and $z$ directions as
  shown in panel (c).}
\label{fig:domain_duct}
\end{figure}

A schematic of the flow in a square duct is presented in Fig.~\ref{fig:domain_duct} to show the physical domain and the computational domain. A two-dimensional simulation is performed, since the flow is fully developed along the stream-wise direction. In addition, the computational domain only covers a quarter of the cross-section as shown in Fig.~\ref{fig:domain_duct}b due to the symmetry of the flow along $y$ and $z$ directions. All lengths are normalized by the height of the computational domain $h=0.5D$, where $D$ is the height of the duct. The Reynolds number $Re$ is based on the height of the computational domain $h$ and bulk velocity $U_b$. Launder-Gibson Reynolds stress
transport model~\cite{gibson78ground} is used for the RANS simulations of both the training
flow and the test flow. DNS data at $Re=2200$ and $3500$ are obtained from Pinelli
et al.~\cite{pinelli10reynolds}. Experimental data at $Re=125000$ is obtained from Gessner and Emery~\cite{gessner81numerical}.

\begin{figure}[htbp]
\centering
\includegraphics[width=0.75\textwidth]{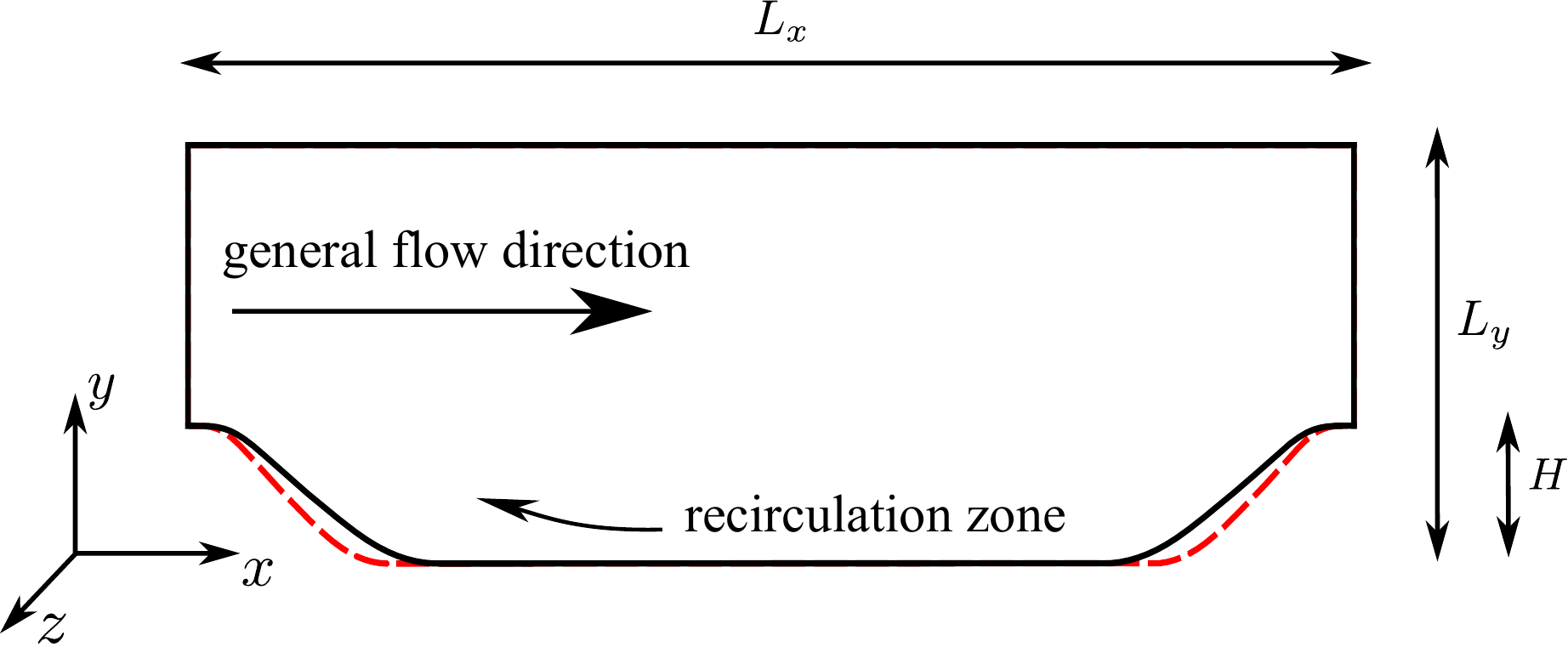}
\caption{Computational domain for the flow over periodic hills. The solid line indicates the configuration of the test flow, and the dashed line indicates the configuration of the training flow. The hill width of the training flow is 0.8 of the hill width of the test flow. The $x$, $y$ and $z$ coordinates
  are aligned with streamwise, wall-normal, and spanwise, respectively.}
\label{fig:domain_pehill}
\end{figure}

Another training-prediction case with the flows over periodic hills is shown in Fig.~\ref{fig:domain_pehill}. The test flow is the flow over periodic hills at $Re=5600$~\cite{breuer09flow}.
The geometry of the computational domain of the test flow is shown in Fig.~\ref{fig:domain_pehill}. The training flow has a steeper hill profile indicated by the dashed line in Fig.~\ref{fig:domain_pehill} . The Reynolds number 
$Re$ is based on the crest height $H$ and the bulk velocity $U_b$ at the crest. Periodic boundary conditions are applied in the streamwise ($x$) direction, 
and non-slip boundary conditions are applied at the walls. The baseline RANS simulations used Launder-Sharma $k$-$\varepsilon$ model~\cite{launder74application}.

All the RANS simulations are performed in an open-source CFD platform OpenFOAM, using a
built-in steady-state incompressible flow solver \texttt{simpleFoam}~\citep{weller98tensorial}, in
which the SIMPLE algorithm~\cite{patankar80numerical} is used. In the RANS simulations, the $y^+$ of the first cell center is kept less than 1 and thus no wall model is applied.

The training-prediction cases in the \textit{a posteriori} test are summarized in Table.~\ref{tab:scenarios}. The case 1 is investigated to show the Reynolds number extrapolation for which DNS data is available to examine the prediction performance in details. The case 2 is chosen to demonstrate the capability of the proposed framework at higher Reynolds number where only experimental data is available. The case 3 is studied to demonstrate the prediction performance for which the training flow and the test flow have different geometry configurations.
\begin{table}[htbp] 	
  \centering
  \caption{The training-prediction scenarios in $\textit{a posteriori}$ test.
}
\label{tab:scenarios}
\begin{tabular}{P{2.0cm} | P{6.0cm}  P{6.0cm} }	
  \hline
 Cases  & Training set & Test set \\ 
  \hline
  1  & Flow in a square duct at $Re=2200$~\cite{pinelli10reynolds} & Flow in a square duct at $Re=3500$~\cite{pinelli10reynolds} \\  
  \hline	
  2  & Flow in a square duct at $Re=2200$~\cite{pinelli10reynolds} & Flow in a square duct at $Re=125000$~\cite{gessner81numerical} \\  
  \hline	
  3  & Flow over periodic hills at $Re=5600$ (steeper hill profile) & Flow over periodic hills at $Re=5600$~\cite{breuer09flow} \\  
  \hline					 								
\end{tabular}
\flushleft
\end{table}

\subsection{Propagation of DNS Reynolds stresses}
In this test, the DNS Reynolds stresses are used to illustrate the merit of physics-based implicit treatment. Three types of turbulence models are compared: Reynolds stress models (RSM) with explicit treatment, linear eddy viscosity models and RSM with implicit treatment. The \textit{explicit treatment} means that the modeled Reynolds stress is directly substituted into the RANS equations to solve for mean velocity as an explicit term. The dependence of Reynolds stress upon the strain rate can still be taken into account by updating the modeling of Reynolds stress during the time stepping (or iterations for steady problems). However, merely updating the Reynolds stress explicitly based on the solved mean velocity would not improve the conditioning of the RANS equations, which has been further discussed in~\cite{wu17on}. The \textit{implicit treatment} means that the modeled Reynolds stress implicitly depends on the strain rate through an optimized eddy viscosity. Such an implicit treatment would improve the conditioning of RANS equations since the optimized eddy viscosity has impact upon the coefficient matrix of the discretized RANS equations and thus influences the condition number.

In the test of Reynolds stress models, the unclosed term in momentum equation is substituted with DNS Reynolds stress. In the test of linear eddy viscosity models, the eddy viscosity term is substituted with the optimal eddy viscosity $\nu_t^L$ obtained from DNS data. Specifically, the optimal eddy viscosity $\nu_t^L$ is computed by minimizing the discrepancy between the linear eddy viscosity model and the DNS Reynolds stress data, i.e., $\nu_t^L = \argmin_{\nu_t} || \mathbf{b}^{DNS} - \nu_t \mathbf{S}^{DNS}||$. Compared with the eddy viscosity models, the RSM with implicit treatment further takes into account the non-linear part of Reynolds stress, i.e., $\bm{\tau}=\nu_t^L\mathbf{S}+(\bm{\tau}^{DNS} - \nu_t^L \mathbf{S}^{DNS})$. Two canonical flows, the flow in a square duct and the flow over periodic hills, are studied to compare the performance of these three types of models. This propagation test demonstrates that the linear eddy viscosity model is unreliable within the region where the misalignment between Reynolds stress tensor and strain-rate tensor is not negligible. In the test with explicit treatment of Reynolds stress, the DNS data is employed to demonstrate the problem of the machine learning modeling approach under an ideal scenario. Specifically, this ideal scenario means that the initial mean velocity field is already the same as DNS mean velocity and the modeled Reynolds stress is also the same as DNS Reynolds stress. Such an ideal scenario represents an absolute performance ceiling of any machine-learning-assisted modeling approach with the explicit treatment. Under this ideal scenario, we demonstrate that a small error in DNS Reynolds stress (e.g., sampling error) can lead to a relatively large error in the solved mean velocity field. If the modeling of Reynolds stress is further updated by evaluating the data-driven model with the solved mean velocity, the errors in the modeled Reynolds stress and the solved mean velocity would be further amplified as has been demonstrated for the plane channel flow in~\cite{wu17on}. Therefore, an implicit treatment is needed to improve the conditioning of the machine-learning-assisted RANS modeling.

\subsubsection{Flow in a square duct}
\begin{figure}[htbp]
\centering
\includegraphics[width=0.25\textwidth]{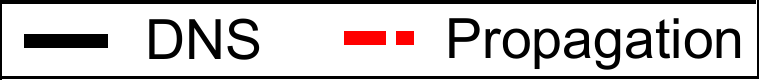}\\
\subfloat[RSM with explicit treatment]{\includegraphics[width=0.44\textwidth]{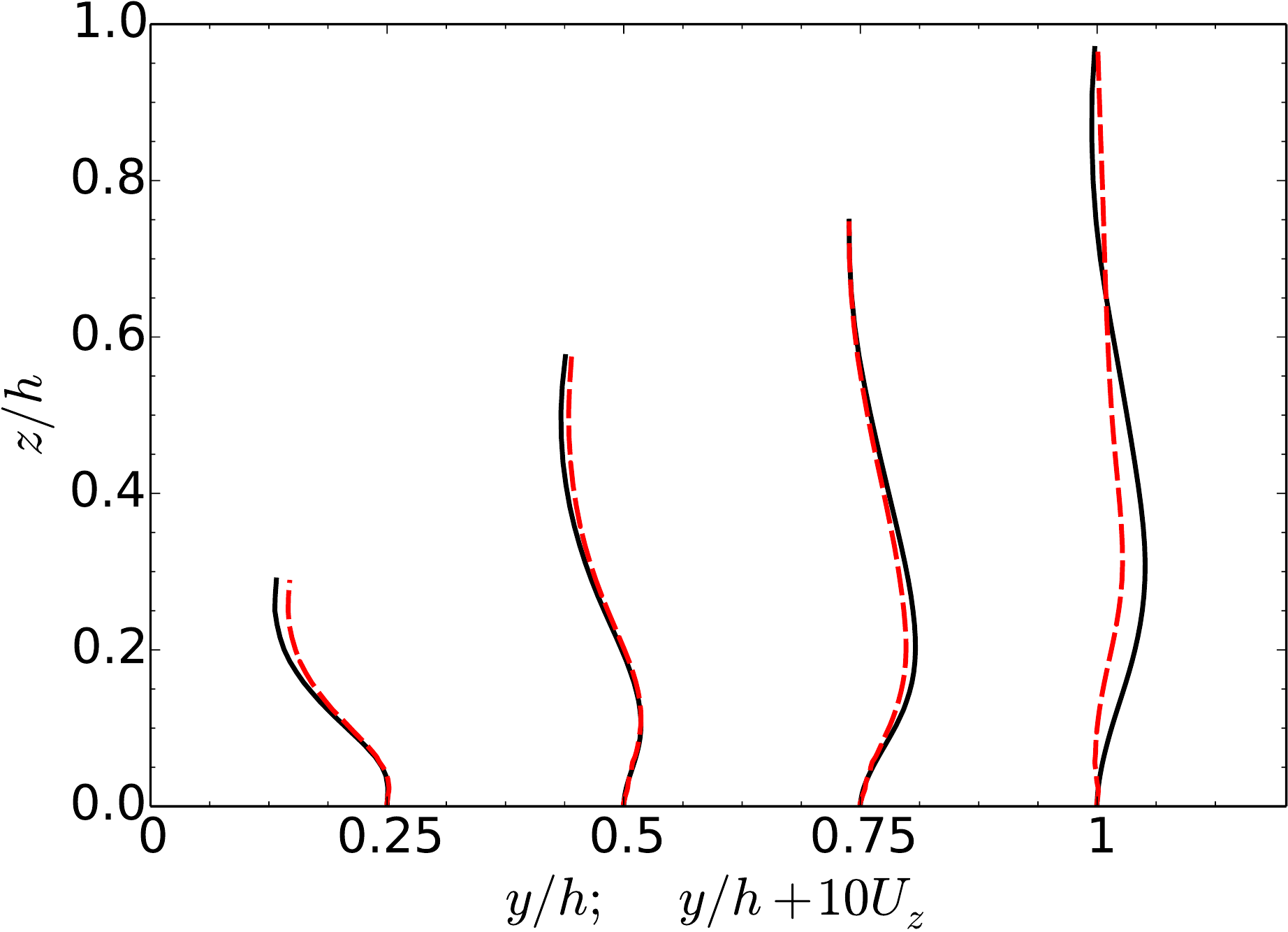}}\hspace{1em}
\subfloat[RSM with implicit treatment]{\includegraphics[width=0.44\textwidth]{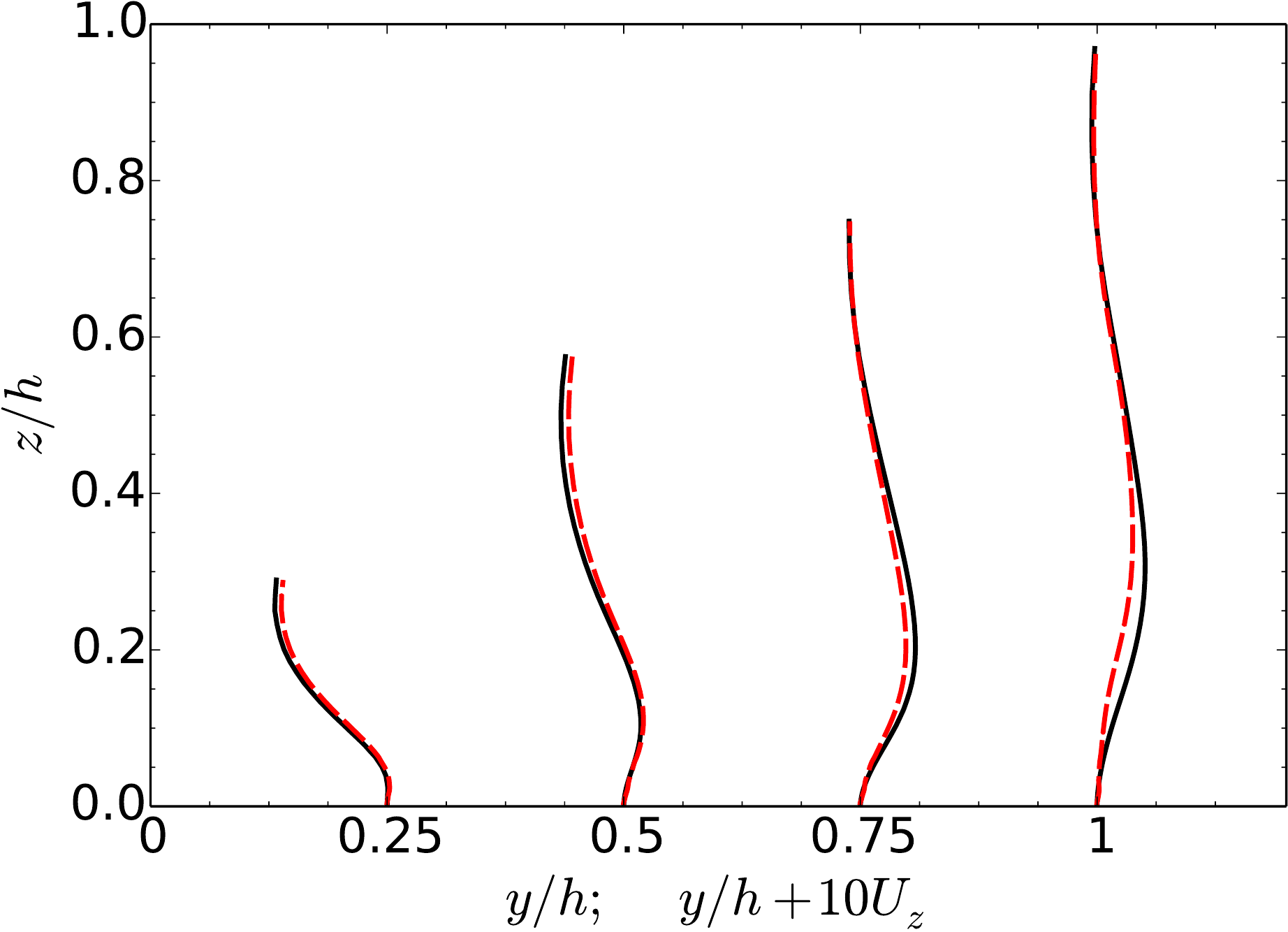}}
\caption{The comparison of secondary flow velocity $U_z$ by using (a) Reynolds stress model with explicit treatment and (b) Reynolds stress model with implicit treatment. The results by using linear eddy viscosity model capture no secondary flow and are thus omitted here.}
\label{fig:Uz-duct-apriori}
\end{figure}

The secondary velocity $U_z$ based on DNS Reynolds stress is shown in Fig.~\ref{fig:Uz-duct-apriori}  by using Reynolds stress models with explicit and implicit treatments. It can be seen in Fig.\ref{fig:Uz-duct-apriori}(a) that the solved secondary velocity $U_z$ agrees well with DNS data within most regions, except along the symmetry plane $y/h=1$. Such good agreement of solved secondary velocity indicates that the Reynolds stress models lead to well-conditioned discretized momentum equations for the flow in a square duct. Similar quality of secondary velocity can be obtained by using RSM with implicit treatment. On the other hand, the linear eddy viscosity models cannot predict the secondary flow due to the limitation in representing the normal stress imbalance, the results of which are omitted here for simplicity.

\subsubsection{Flow over periodic hills}
\begin{figure}[htbp]
\centering
\hspace{2em}\includegraphics[width=0.25\textwidth]{apriori-legend}\\
\subfloat[RSM with explicit treatment]{\includegraphics[width=0.6\textwidth]{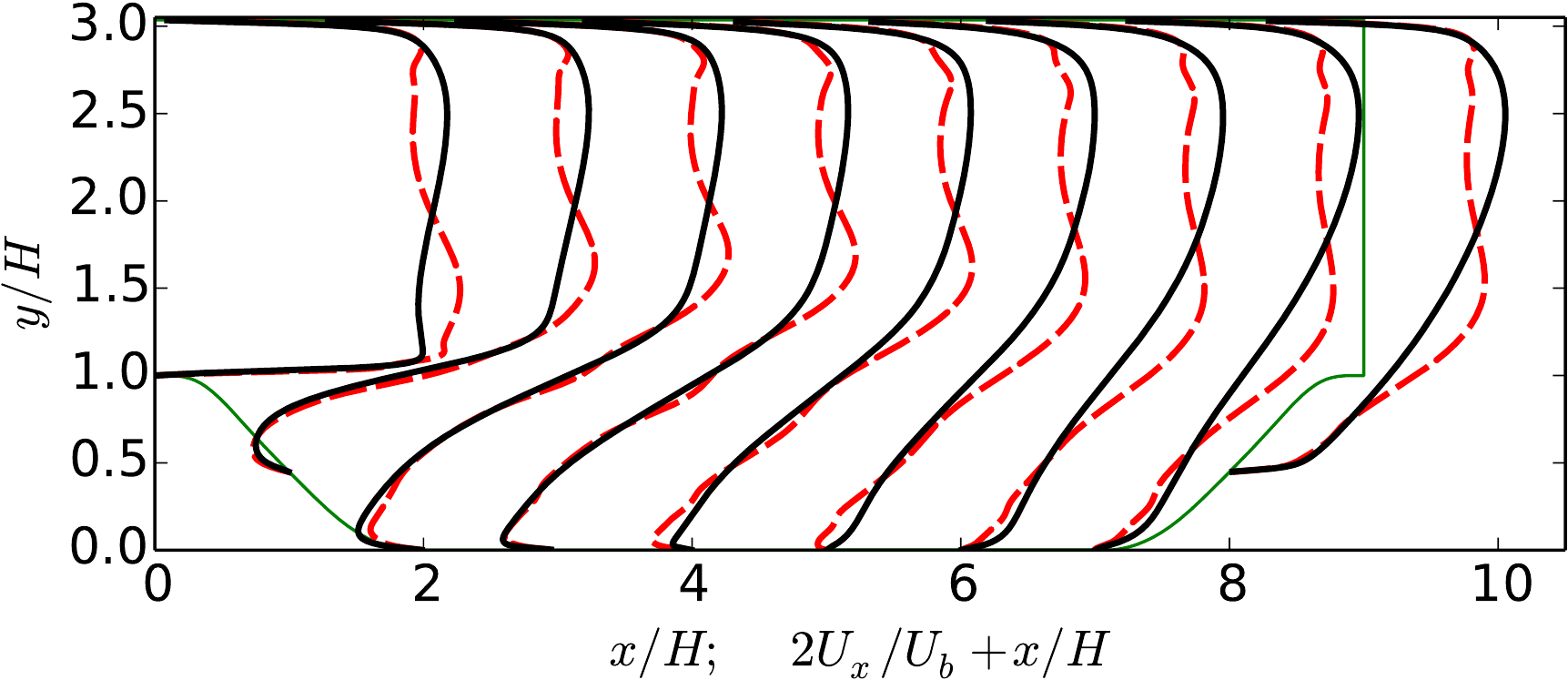}}\\
\subfloat[Eddy viscosity model]{\includegraphics[width=0.6\textwidth]{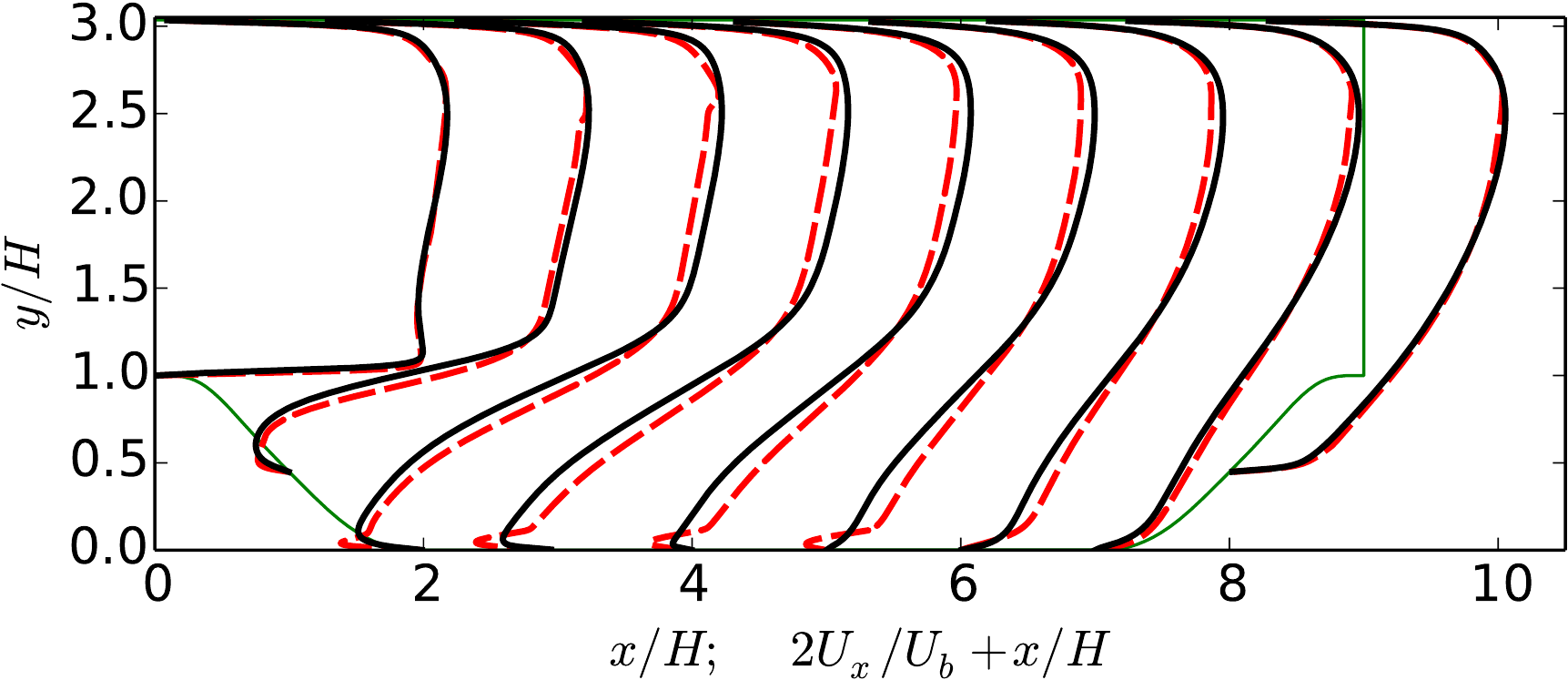}}\\
\subfloat[RSM with implicit treatment]{\includegraphics[width=0.6\textwidth]{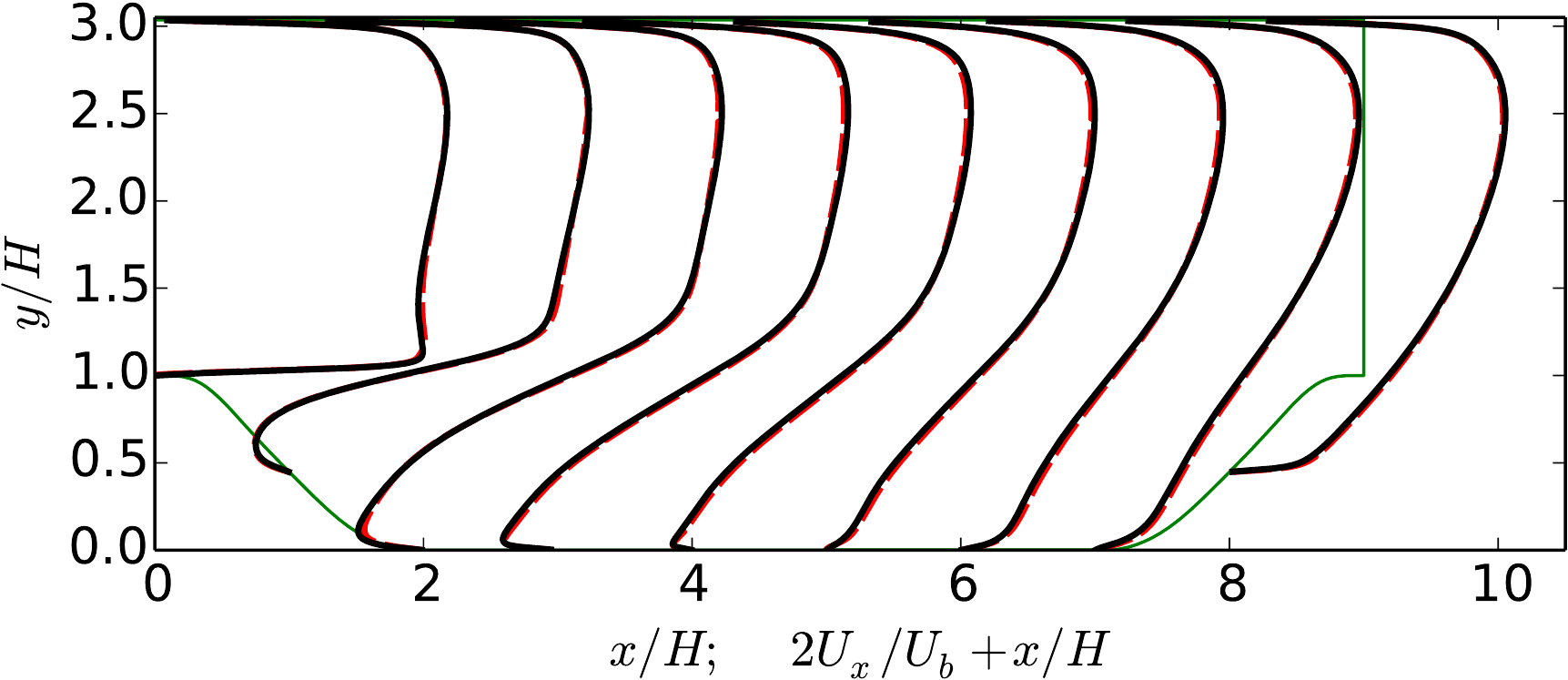}}
\caption{The solved mean velocity field for the flow over periodic hills at $Re=5600$ by using (a) Reynolds stress model with explicit treatment, (b) eddy viscosity model and (c) Reynolds stress model with implicit treatment. The DNS data is utilized as the modeled term to represent the best possible performance among the respective class of models.}
\label{fig:U-tauFoam}
\end{figure}

Although the Reynolds stress models with explicit treatment perform well for the flow in a square duct, they are potentially unreliable since it can lead to pronounced errors in mean velocity even for the turbulent channel flows~\cite{thompson16methodology,wu17on}. In this work, we demonstrate the issue of Reynolds stress models with explicit treatment for the flow with massive separation. Figure~\ref{fig:U-tauFoam}(a) shows that the solved mean velocity field does not agree with the DNS data by using Reynolds stress model with explicit treatment for the flow over periodic hills. It should be noted that the velocity shown in Fig.~\ref{fig:U-tauFoam}(a) is obtained by using DNS Reynolds stress, and not the modeled Reynolds stress. Therefore, the unsatisfactory results in Fig.~\ref{fig:U-tauFoam}(a) indicates the best possible performance of machine-learning-assisted turbulence modeling via directly substituting the modeled Reynolds stress into RANS equations.

\begin{figure}[htbp]
\centering
\includegraphics[width=0.5\textwidth]{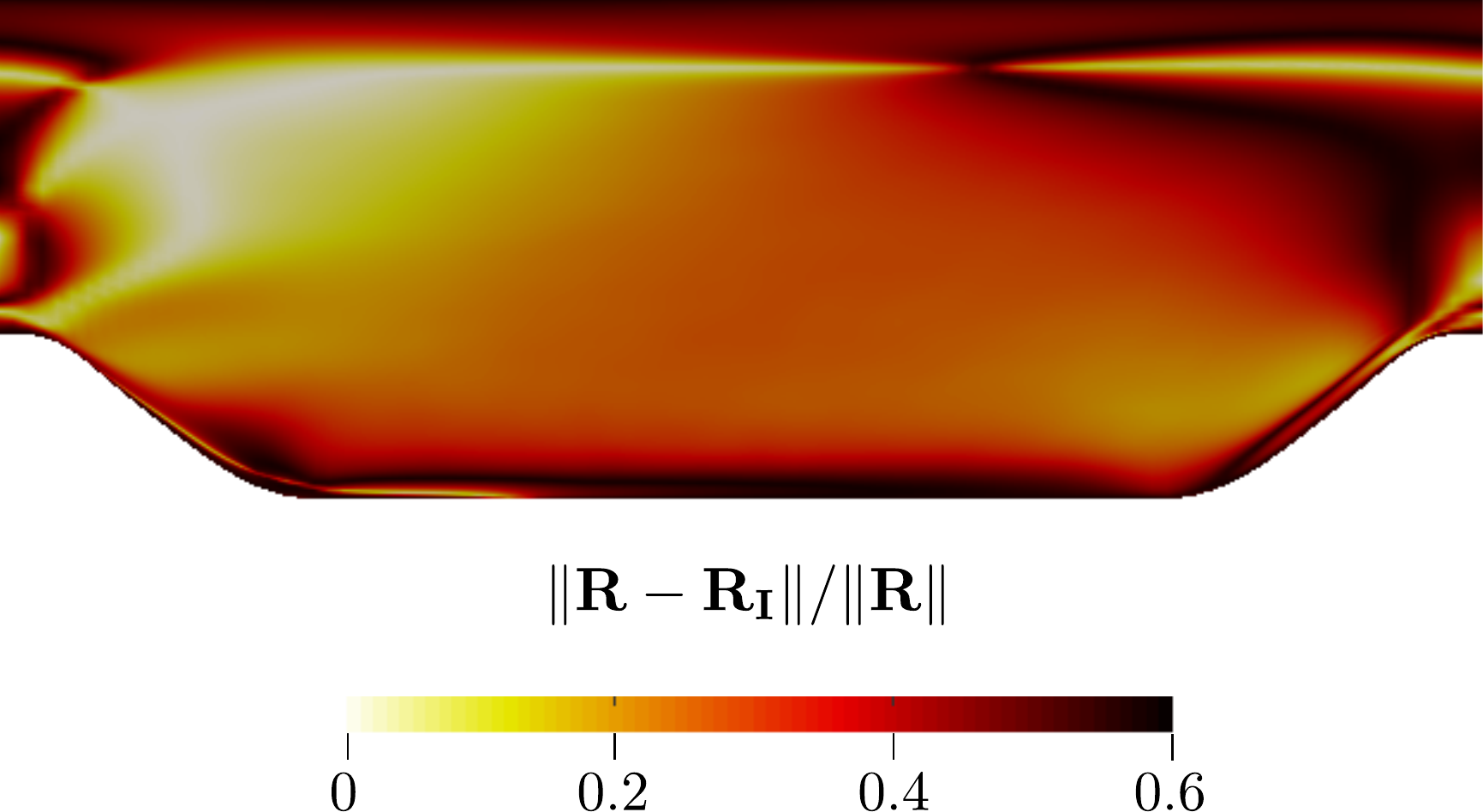}
\caption{An indicator of the deficiency of linear eddy viscosity model due to the misalignment of eigenvectors between Reynolds stress tensor and strain rate tensor.}
\label{fig:R-misalign}
\end{figure}

Better results of mean velocity can be achieved as shown in Fig.~\ref{fig:U-tauFoam}(b) by using linear eddy viscosity models than Reynolds stress models with explicit treatment. However, noticeable differences can still be observed between the solved mean velocity and the DNS data. The main reason is that the misalignment between the Reynolds stress tensor and the strain rate tensor is neglected by using linear eddy viscosity models. Such misalignment can be quantified by the rotation matrix $\mathbf{R}$ from the eigenvectors of Reynolds stress tensor to the ones of strain rate tensor. Figure~\ref{fig:R-misalign} presents an indicator of misalignment calculated based on the deviation of the rotation matrix $\mathbf{R}$ from the identity matrix~$\mathbf{R_I}$.

Unlike the eddy viscosity models, the RSM with implicit treatment take into account the difference between the linear part of Reynolds stress $\bm{\tau}^{L}$ and the true Reynolds stress $\bm{\tau}$. The purpose is to address the misalignment of eigenvectors between Reynolds stress tensor and strain rate tensor. It can be seen in Fig.~\ref{fig:U-tauFoam}(c) that the solved mean velocity field has a much better agreement with DNS data, compared with the results by using the Reynolds stress models with explicit treatment in Fig.~\ref{fig:U-tauFoam}(a) and the eddy viscosity models in Fig.~\ref{fig:U-tauFoam}(b). By using DNS Reynolds stress data as the ideal machine-learning-modeled stress, the propagation test results in Fig.~\ref{fig:U-tauFoam} demonstrate the superiority of RSM with implicit treatment in achieving predicative capability of mean velocity field.

\subsection{A posteriori test}

\subsubsection{Flow in a square duct}
\label{sec:duct}
\begin{figure}[htbp]
\centering
\includegraphics[width=0.3\textwidth]{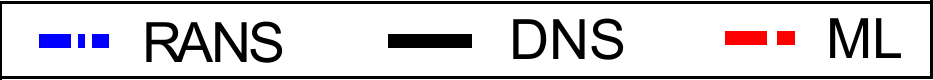} \\
\subfloat[$\bm{\tau}^{\textrm{DNS}}_{yy}$]{\includegraphics[width=0.4\textwidth]{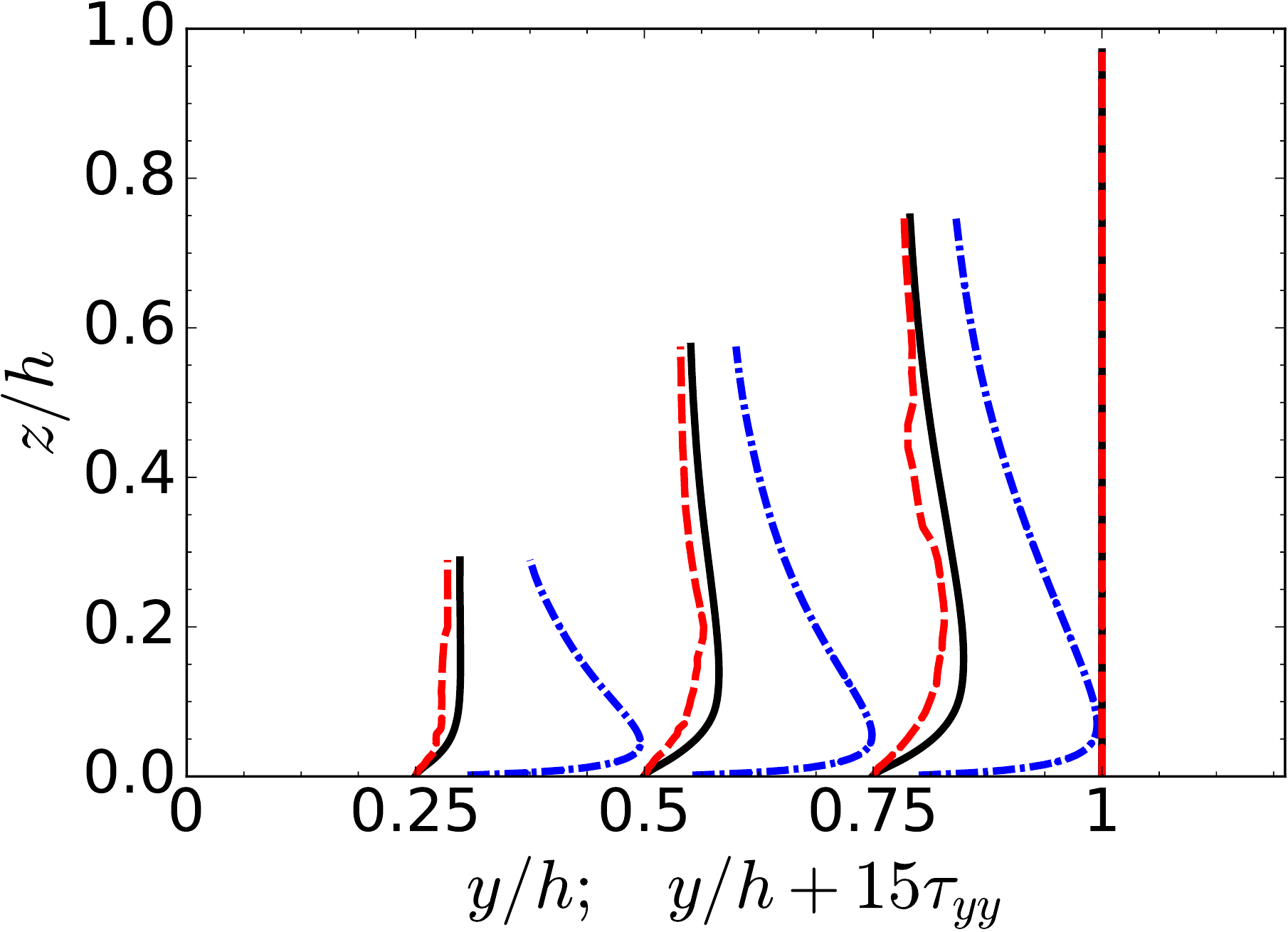}} \hspace{0.5em}
\subfloat[$\bm{\tau}^{\textrm{DNS}}_{zz}$]{\includegraphics[width=0.4\textwidth]{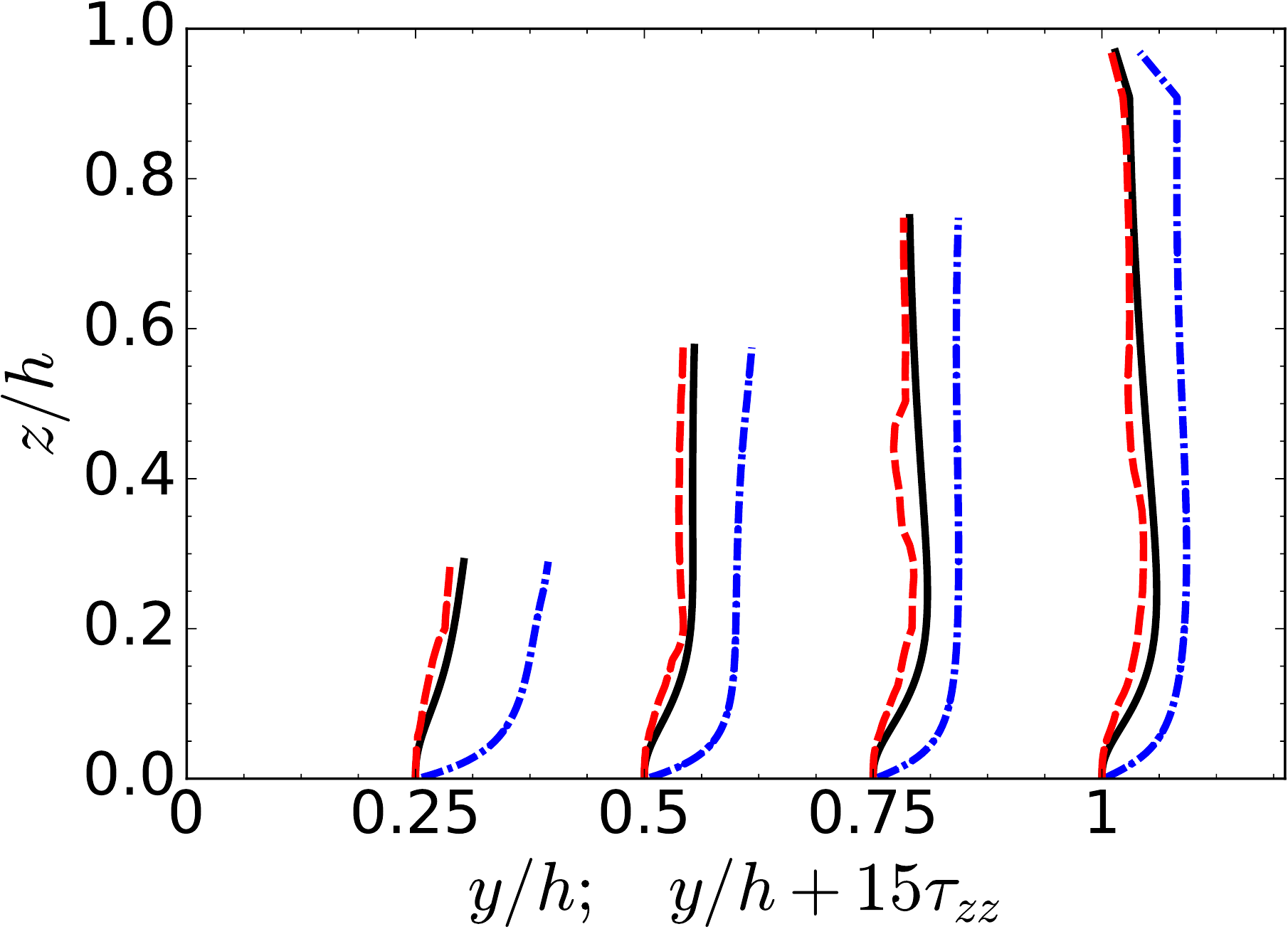}}
\caption{The prediction of \textbf{DNS Reynolds stress components} of the flow in a square duct at Reynolds number $Re=3500$, including (a) $\bm{\tau}^{\textrm{DNS}}_{yy}$ and (b) $\bm{\tau}^{\textrm{DNS}}_{zz}$ at $y/h=0.25, 0.5, 0.75$ and 1. The training flow is at Reynolds number $Re=2200$.}
\label{fig:Tau_duct}
\end{figure}

In the first case, the random forest is trained by using the flow in a square duct at Reynolds number $Re=2200$. The flow at Reynolds number $Re=3500$ is used as the test flow. It should be noted that the DNS data at $Re=3500$ is only used to evaluate the machine learning prediction, and not being used for training the machine learning model. The baseline RANS indicates the results obtained from standard RANS simulations. The machine learning results are denoted by ML in the legends of figures for simplicity. It can be seen in Fig.~\ref{fig:Tau_duct} that the baseline RANS simulated normal components of Reynolds stress qualitatively captures the imbalance between $\bm{\tau}_{yy}$ and $\bm{\tau}_{zz}$. However, the simulated normal stress imbalance is noticeably greater than the DNS data, especially around the near wall region. Such greater normal stress imbalance between $\bm{\tau}_{yy}$ and $\bm{\tau}_{zz}$ explains the stronger secondary flow of baseline RANS simulation. Compared with the baseline RANS simulated stress components, the machine-learning-predicted normal stress components $\bm{\tau}_{yy}$ and $\bm{\tau}_{zz}$ demonstrate a much better agreement with the DNS data in Fig.~\ref{fig:Tau_duct}.

\begin{figure}[htbp]
\centering
\includegraphics[width=0.3\textwidth]{Tau-legend.pdf} \\
\subfloat[$\bm{\tau}^{L}_{yy}$]{\includegraphics[width=0.4\textwidth]{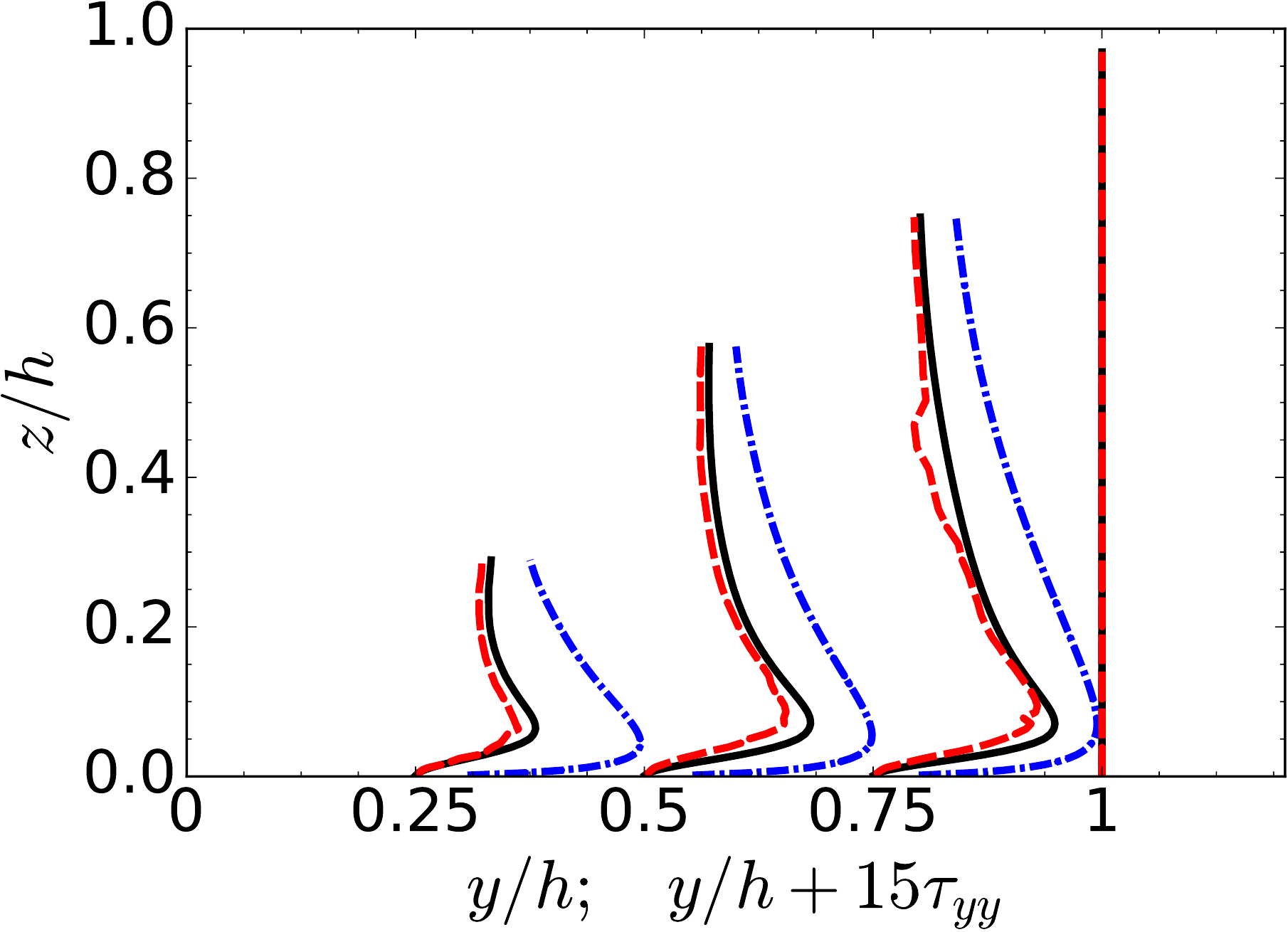}} \hspace{0.5em}
\subfloat[$\bm{\tau}^{L}_{zz}$]{\includegraphics[width=0.4\textwidth]{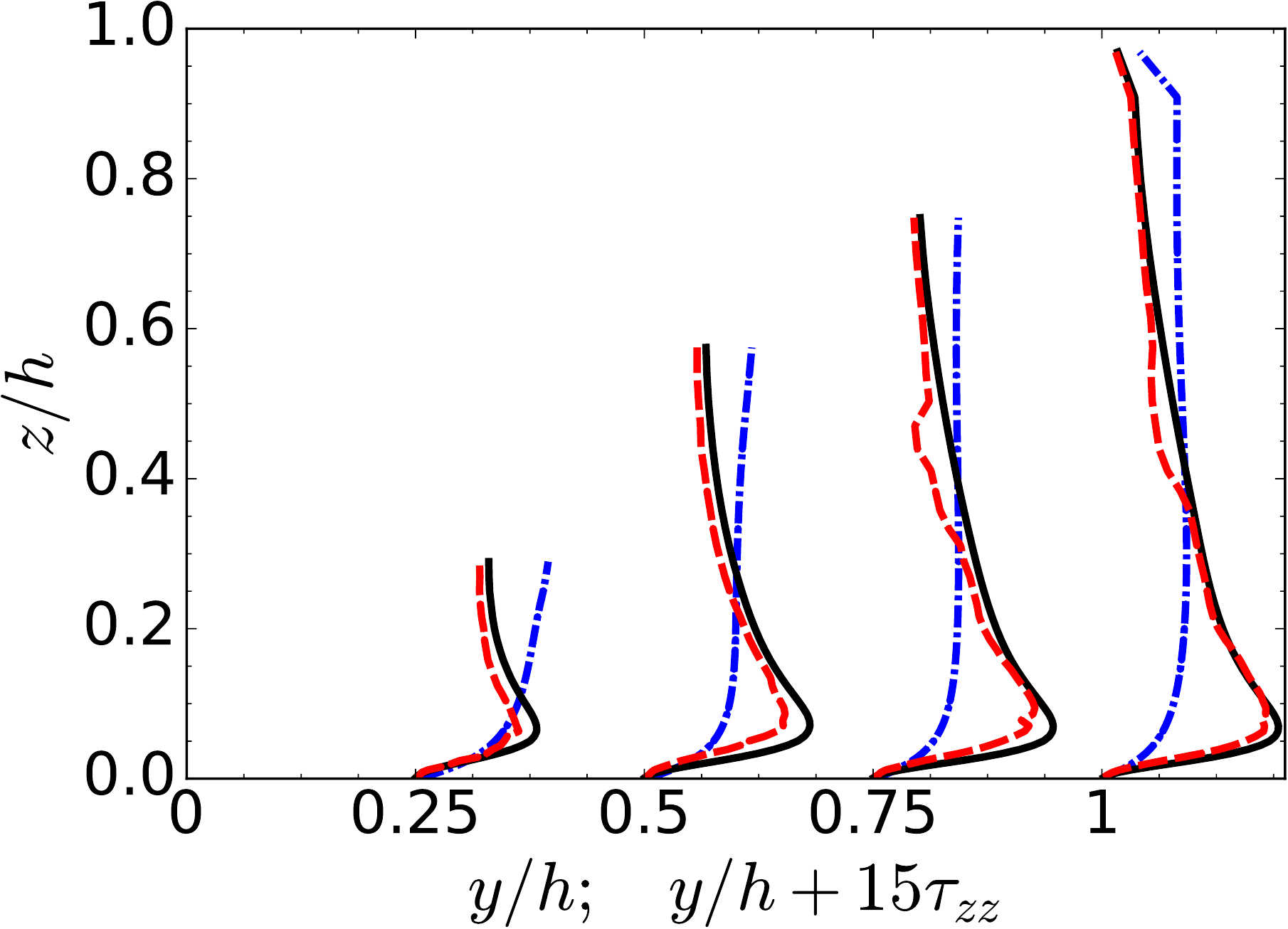}}
\caption{The prediction of \textbf{linear part of Reynolds stress components} of the flow in a square duct at Reynolds number $Re=3500$, including (a) $\bm{\tau}^{L}_{yy}$ and (b) $\bm{\tau}^{L}_{zz}$ at $y/h=0.25, 0.5, 0.75$ and 1. The training flow is at Reynolds number $Re=2200$.}
\label{fig:TauLS_duct}
\end{figure}

The linear part of Reynolds stress $\bm{\tau}^{L}$ is predicted and presented in Fig.~\ref{fig:TauLS_duct}. It can be seen that the normal components $\bm{\tau}^{L}_{yy}$ and $\bm{\tau}^{L}_{zz}$ of linear part of Reynolds stress are similar to each other. The main reason is that the linear part of Reynolds stress is obtained by projecting the DNS Reynolds stress onto the strain rate tensor and neglecting the nonlinear parts of the DNS Reynolds stress. Therefore, the linear part of Reynolds stress follows the eddy viscosity assumption and thus would have no normal stress imbalance. It can be seen in Fig.~\ref{fig:TauLS_duct} that the machine-learning-predicted linear part of Reynolds stress shows a good agreement with the linear part of Reynolds stress obtained from DNS data.

\begin{figure}[htbp]
\centering
\includegraphics[width=0.4\textwidth]{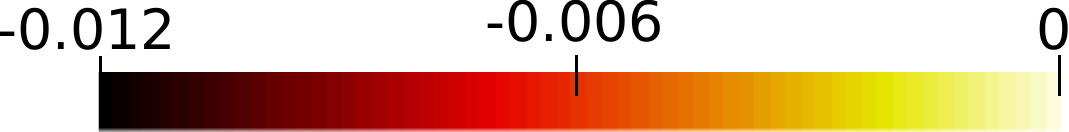}\\
\subfloat[RANS]{\includegraphics[width=0.3\textwidth]{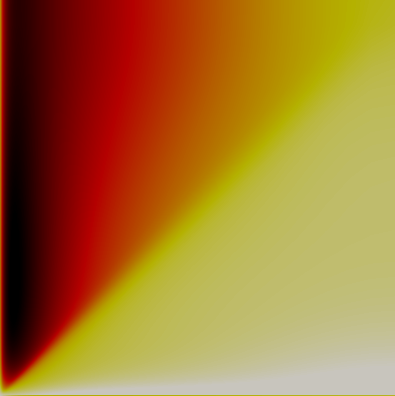}} \hspace{0.2em}
\subfloat[DNS]{\includegraphics[width=0.3\textwidth]{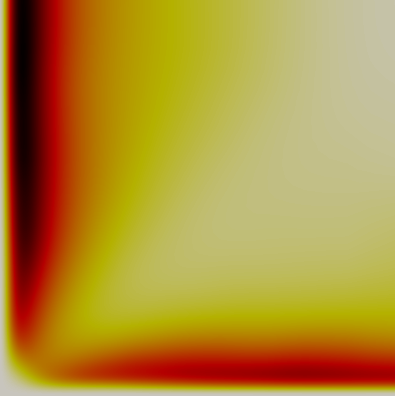}} \hspace{0.2em}
\subfloat[Machine Learning]{\includegraphics[width=0.3\textwidth]{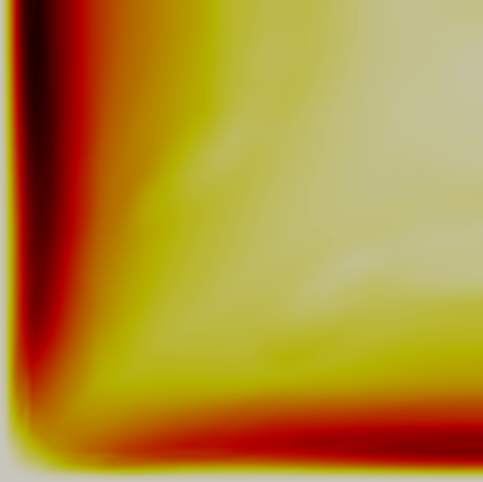}} \hspace{0.2em}
\caption{The non-linear part of Reynolds stress $\bm{\tau}_{yy}^{\perp}$ in a square duct at Reynolds number $Re=3500$, including (a) RANS simulated results, (b) DNS data and (c) prediction of RSM with implicit treatment. The training flow is at Reynolds number $Re=2200$. The color of the contour denotes the value of the stress component, and the light color here indicates small magnitude.}
\label{fig:Tau_perp_contour}
\end{figure}

With both the satisfactory machine learning prediction of Reynolds stress in Fig.~\ref{fig:Tau_duct} and its linear part in Fig.~\ref{fig:TauLS_duct}, it can be expected that the nonlinear term $\bm{\tau}^{\perp}=\bm{\tau}-\bm{\tau}^{L}$ would have a good agreement with the nonlinear part of the DNS Reynolds stress. The comparison of the non-linear Reynolds stress term is presented in Fig.~\ref{fig:Tau_perp_contour}. It can be seen that the noticeable negative stress $\bm{\tau}_{yy}^{\perp}$ can be seen near both the side wall and the bottom wall of the duct based on the DNS data. However, the negative non-linear stress is over-predicted near the side wall and under-predicted near the bottom wall for the RANS simulation. Compared with the RANS results, the machine learning prediction in Fig.~\ref{fig:Tau_perp_contour} demonstrates a much better agreement with the pattern of DNS data.

\begin{figure}[htbp]
\centering
\hspace{2em}\includegraphics[width=0.2\textwidth]{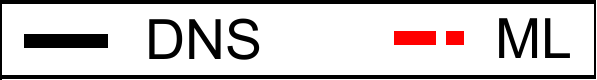} \\
\includegraphics[width=0.5\textwidth]{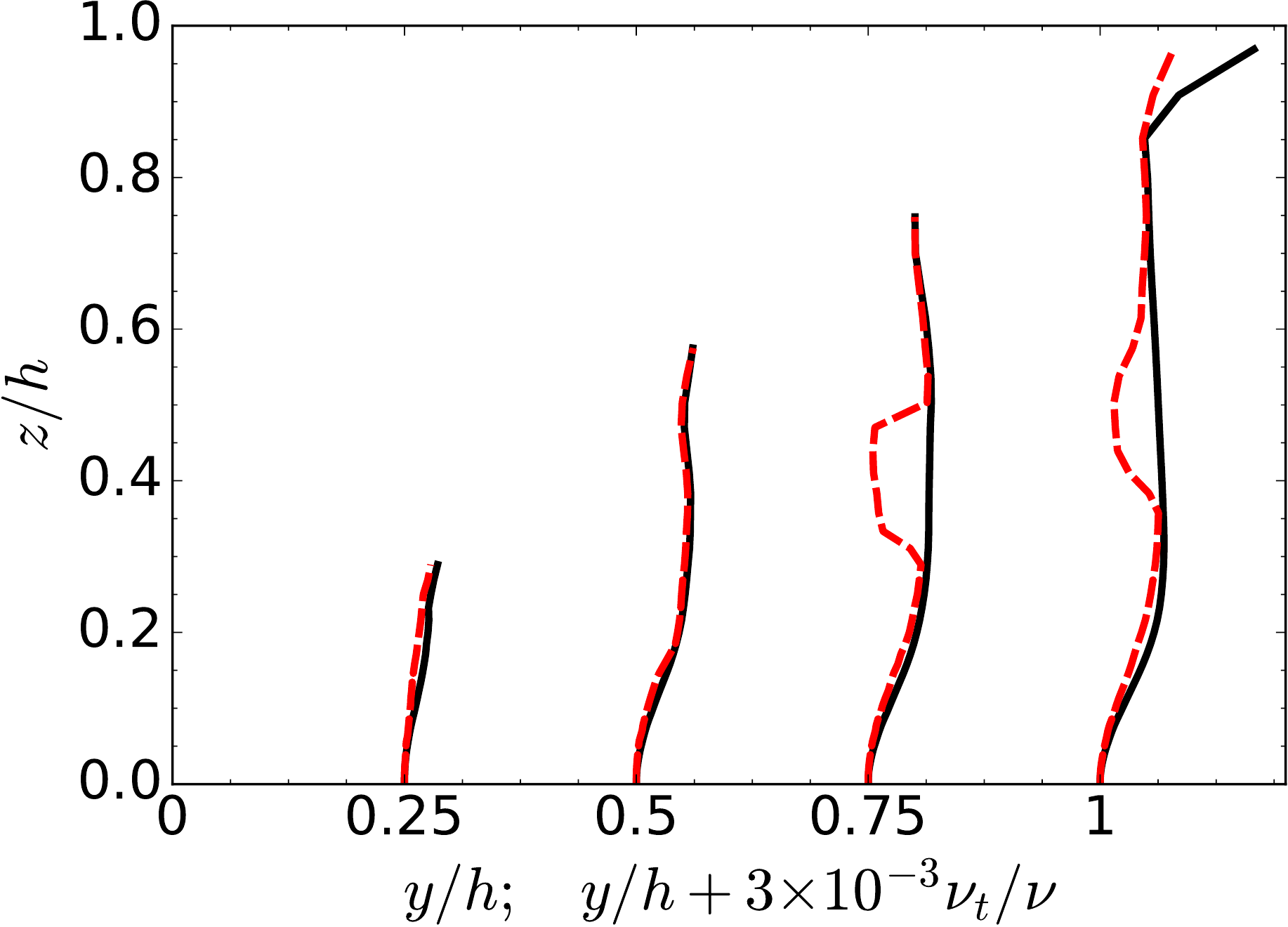}
\caption{The optimal eddy viscosity $\nu_t^L$ of the flow in a square duct at Reynolds number $Re=3500$ at $y/h=0.25, 0.5, 0.75$ and 1. The training flow is at Reynolds number $Re=2200$.}
\label{fig:nutLS_duct}
\end{figure}

\begin{figure}[htbp]
\centering
\subfloat[$U_y$]{\includegraphics[width=0.45\textwidth]{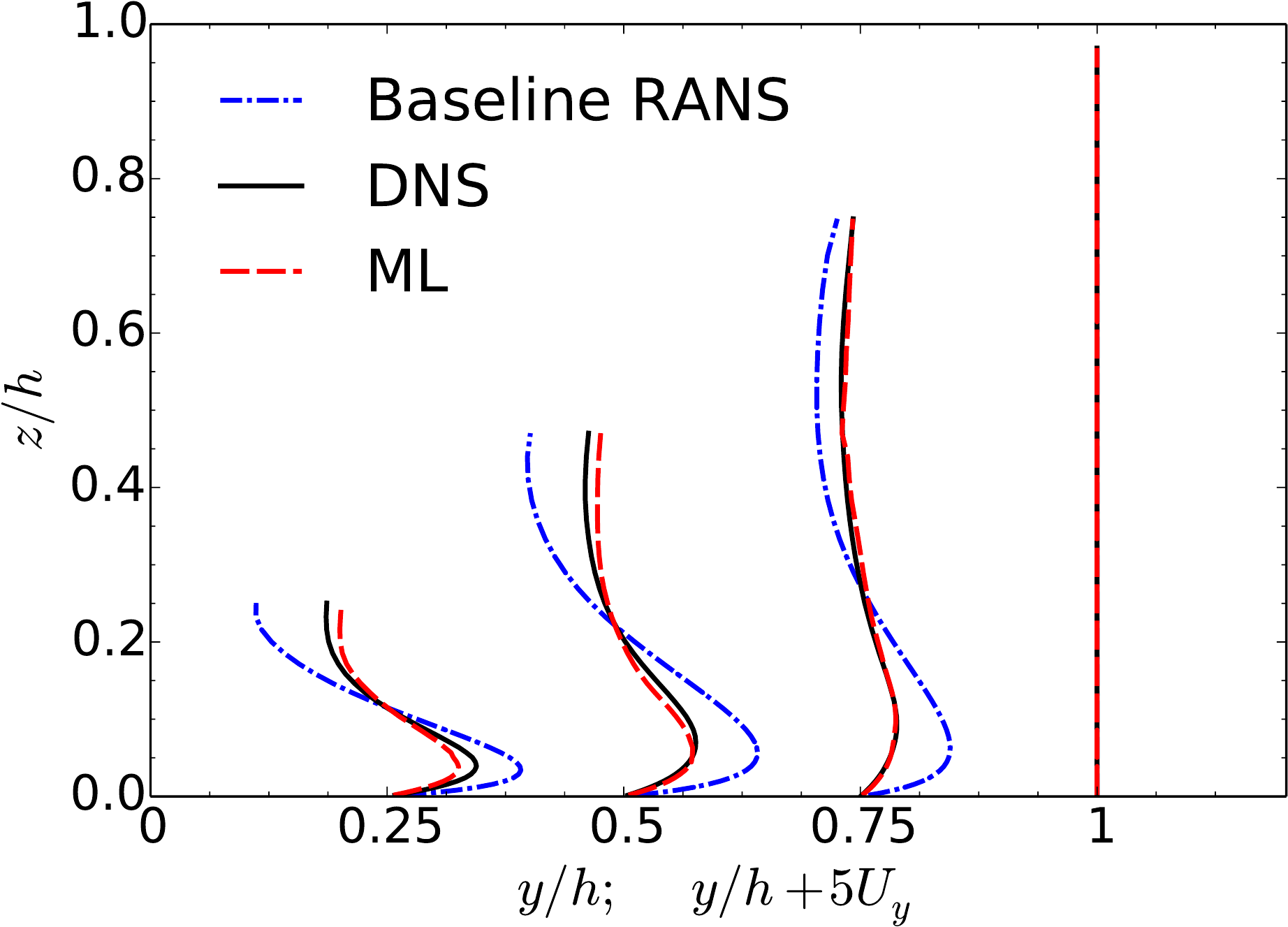}} \hspace{0.5em}
\subfloat[$U_z$]{\includegraphics[width=0.45\textwidth]{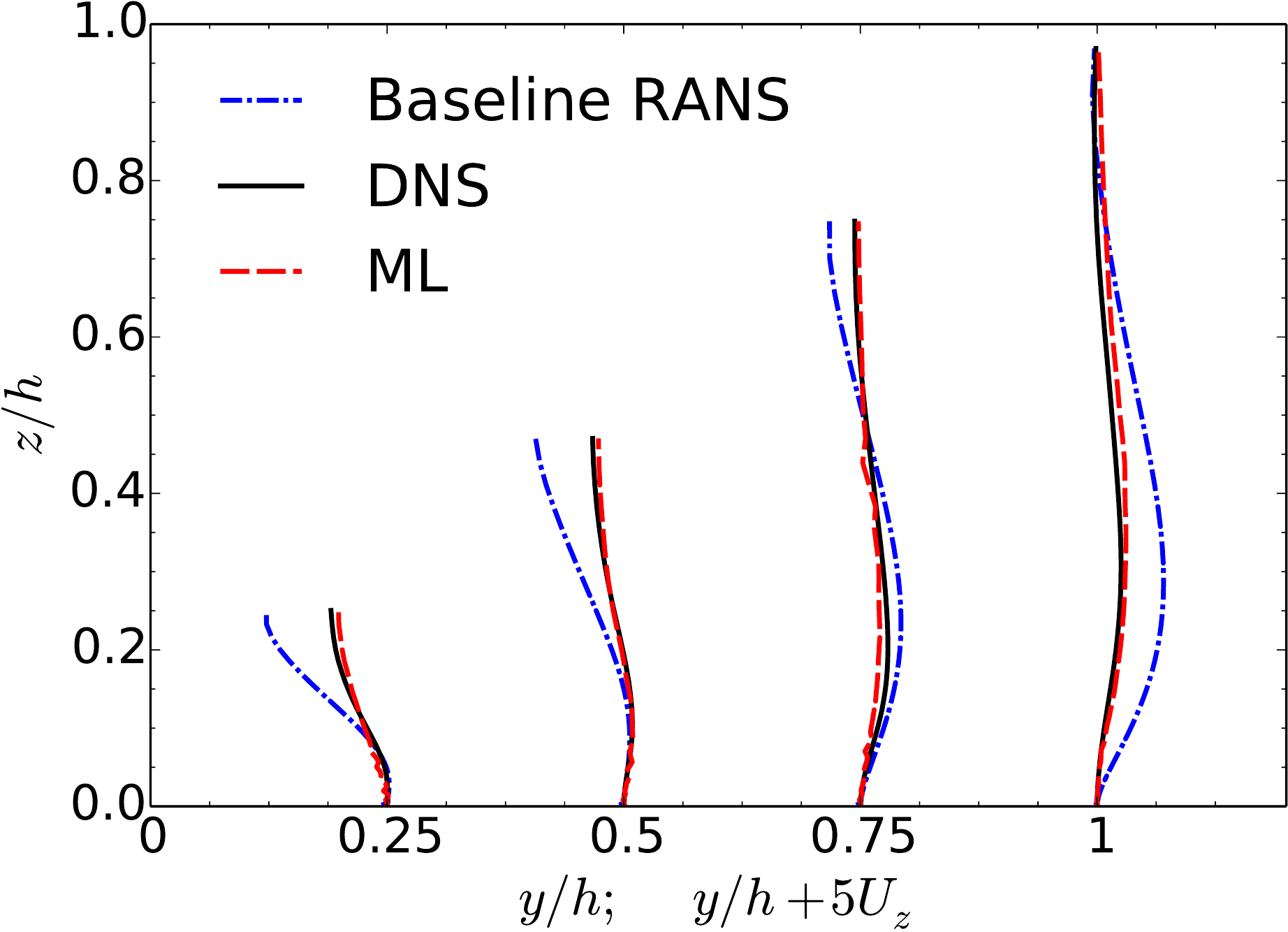}}
\caption{The secondary flow in a square duct at Reynolds number $Re=3500$ predicted by RSM with implicit treatment at $y/h=0.25, 0.5, 0.75$ and 1. The training flow is at Reynolds number $Re=2200$.}
\label{fig:U_duct}
\end{figure}

In addition to the nonlinear part of Reynolds stress, the optimal eddy viscosity $\nu_t^{L}$ is also needed in solving for the mean velocity. It can be seen in Fig.~\ref{fig:nutLS_duct} that the eddy viscosity is close to zero in the near-wall region and increases towards to the diagonal of the duct. The machine-learning-predicted eddy viscosity agrees well with the DNS eddy viscosity at most regions in Fig.~\ref{fig:nutLS_duct}. It should be noted that a few noticeable differences can be observed between the machine-learning-predicted eddy viscosity and the DNS data, e.g., along $y/h=0.75$ and $y/h=1$. However, the velocity gradient is relatively small within these regions and thus such difference has little influence in solving for mean velocity field as demonstrated in Fig.~\ref{fig:U_duct}.

\begin{figure}[htbp]
\centering
\subfloat[RANS]{\includegraphics[width=0.33\textwidth]{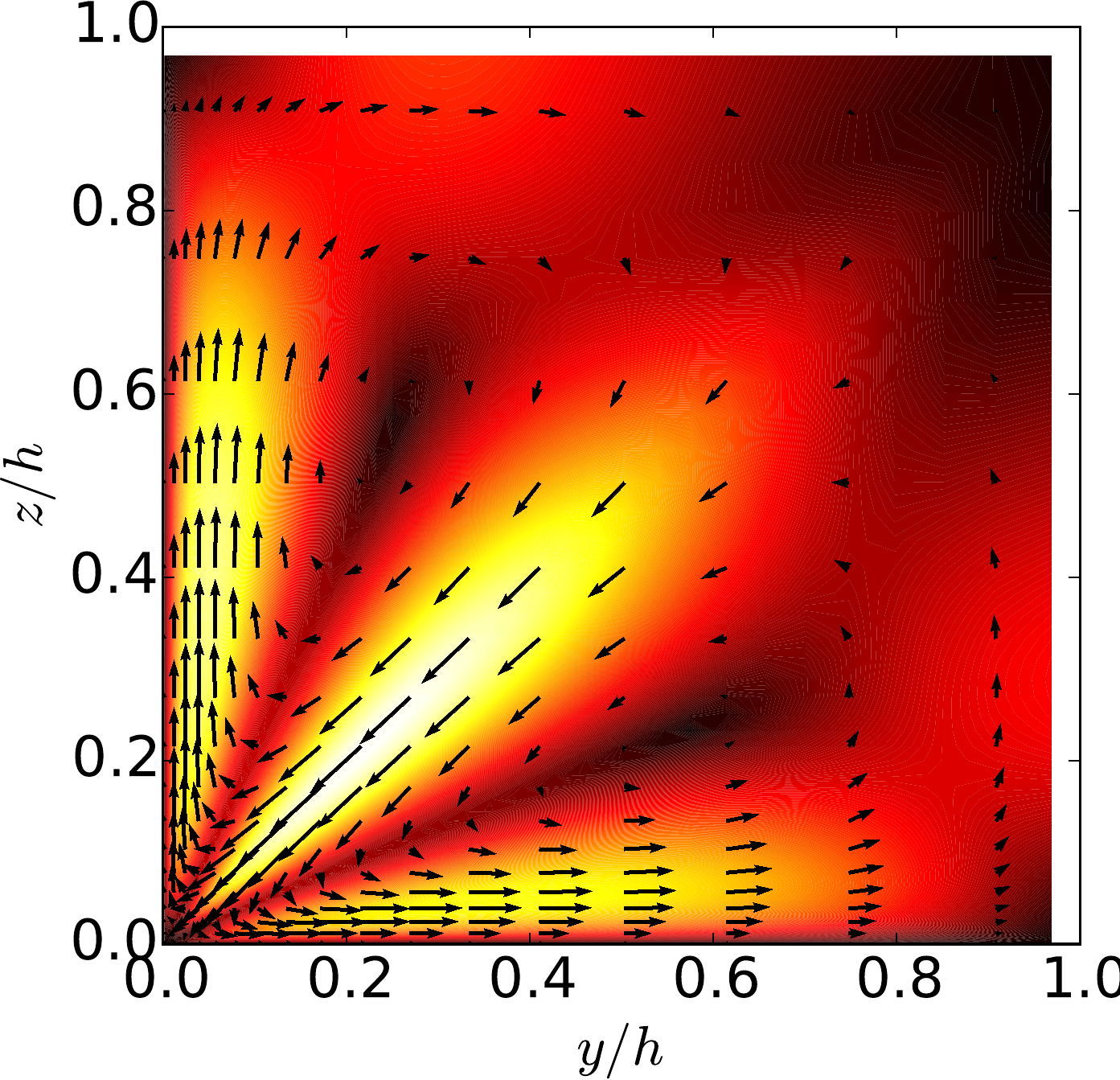}}
\subfloat[DNS]{\includegraphics[width=0.33\textwidth]{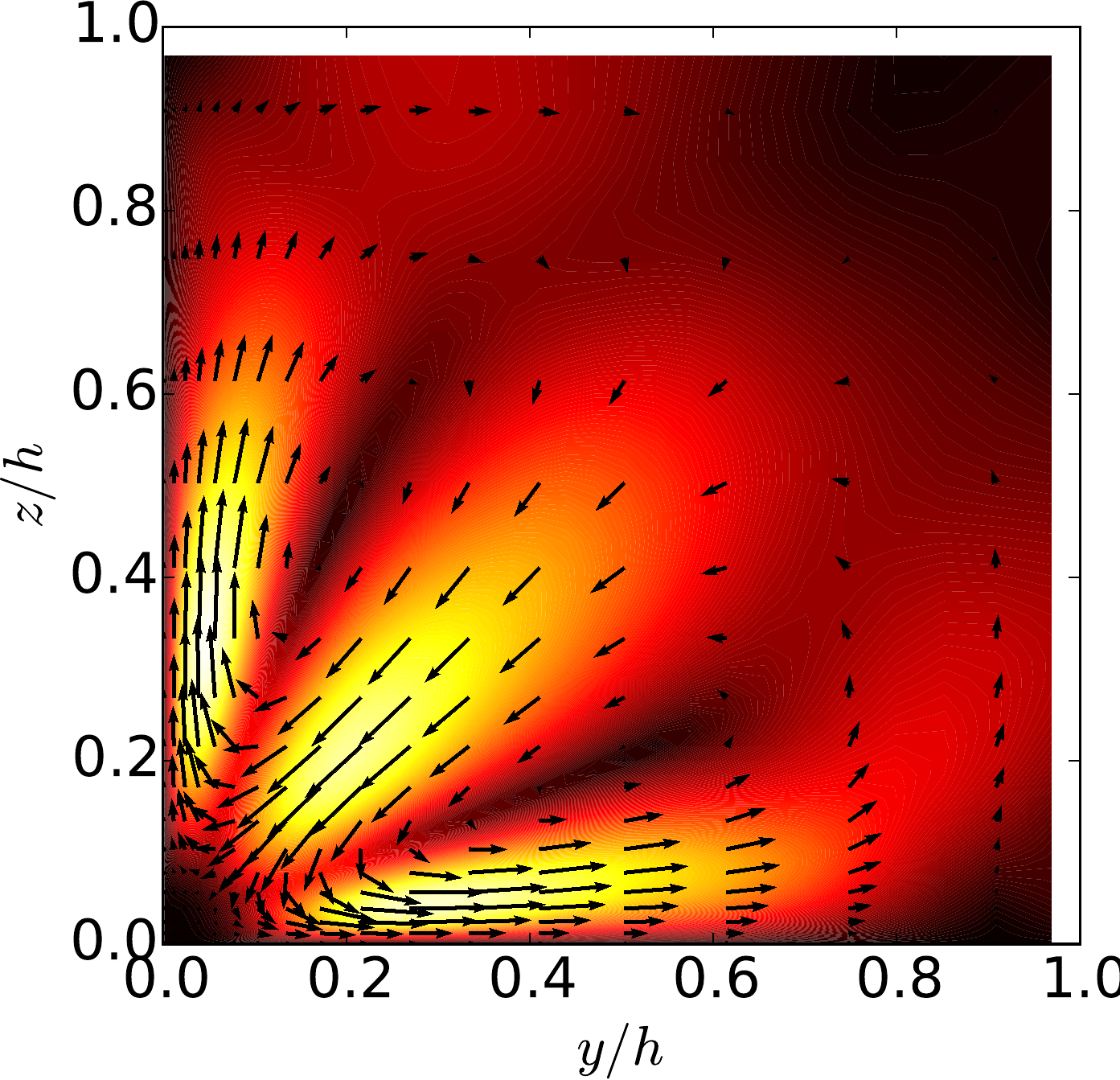}}
\subfloat[Machine Learning]{\includegraphics[width=0.33\textwidth]{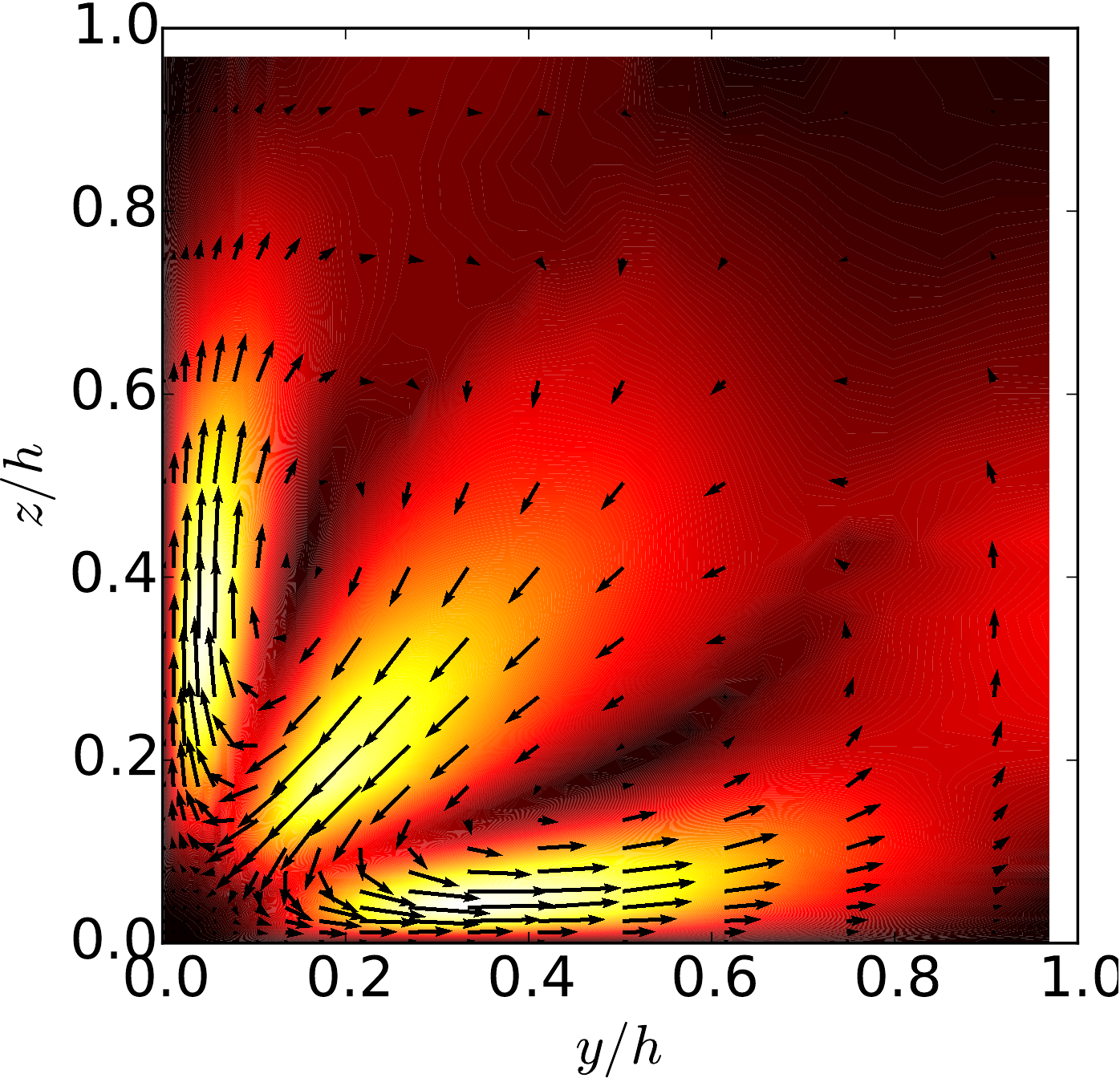}}
\caption{The secondary flow fields in a square duct at Reynolds number $Re=3500$, including (a) RANS simulated results, (b) DNS data and (c) prediction of RSM with implicit treatment. The training flow is at Reynolds number $Re=2200$. The color of the contour denotes the magnitude of the secondary flow. The light color here indicates large magnitude.}
\label{fig:U_duct_contour}
\end{figure}

By substituting the machine-learning-predicted eddy viscosity and nonlinear part of Reynolds stress into RANS equations, the mean velocity field is solved and presented in Fig.~\ref{fig:U_duct_contour}. It can be seen in Fig.~\ref{fig:U_duct_contour}a that the RANS simulated secondary flow penetrates too much toward the left bottom region (corner region between the perpendicular walls). The machine learning predicted secondary flow in Fig.~\ref{fig:U_duct_contour}c demonstrates a better agreement with DNS data in this corner region. In addition, the shape and location of the secondary vortex are better predicted by our machine learning framework as shown in Fig.~\ref{fig:U_duct_contour}c. 

Four profiles of secondary flow are presented in Fig.~\ref{fig:U_duct} for a more quantitative comparison of the secondary flow prediction. It can be seen that the baseline RANS simulated mean velocity field overestimates the magnitude of the secondary flow, especially around the corner region. On the other hand, the mean velocity based on the RSM with implicit treatment shows a much better agreement with the DNS data. However, it should be noted that the test flow is at Reynolds number $Re=3500$, close to that of the training flow ($Re=2200$). Therefore, the satisfactory predictive capability as demonstrated in Fig.~\ref{fig:U_duct} does not necessarily guarantee similar performance at a higher Reynolds number. 

\begin{figure}[htbp]
\centering
\includegraphics[width=0.55\textwidth]{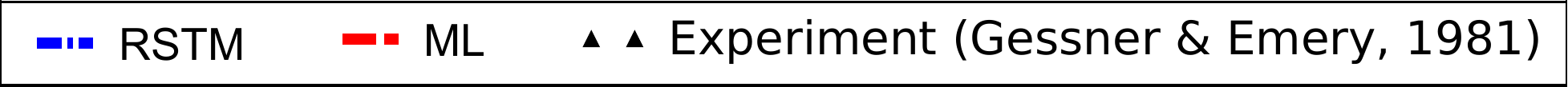}\\
\subfloat[Along vertical axis of symmetry]{\includegraphics[width=0.45\textwidth]{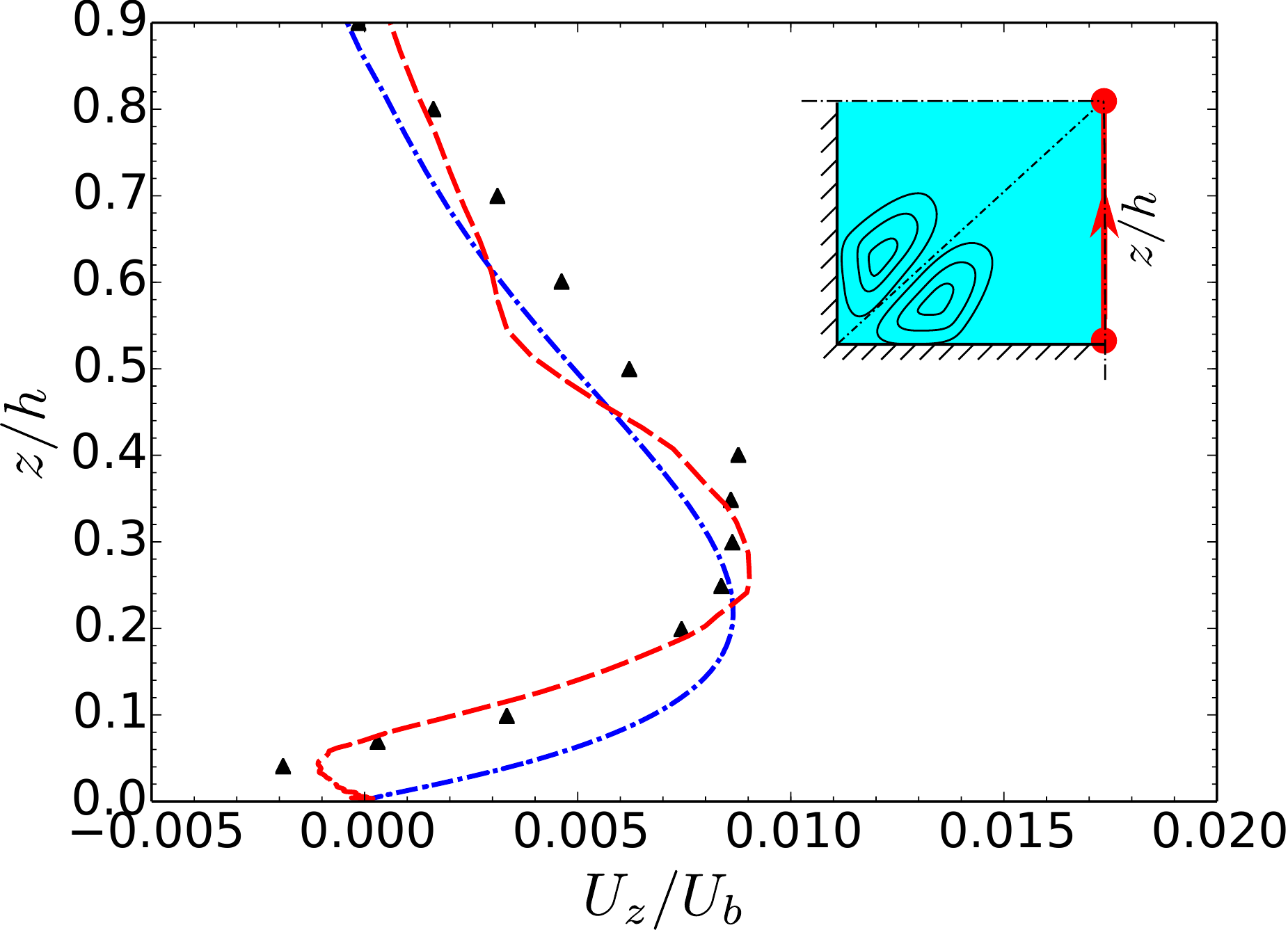}}
\subfloat[Along diagonal]{\includegraphics[width=0.45\textwidth]{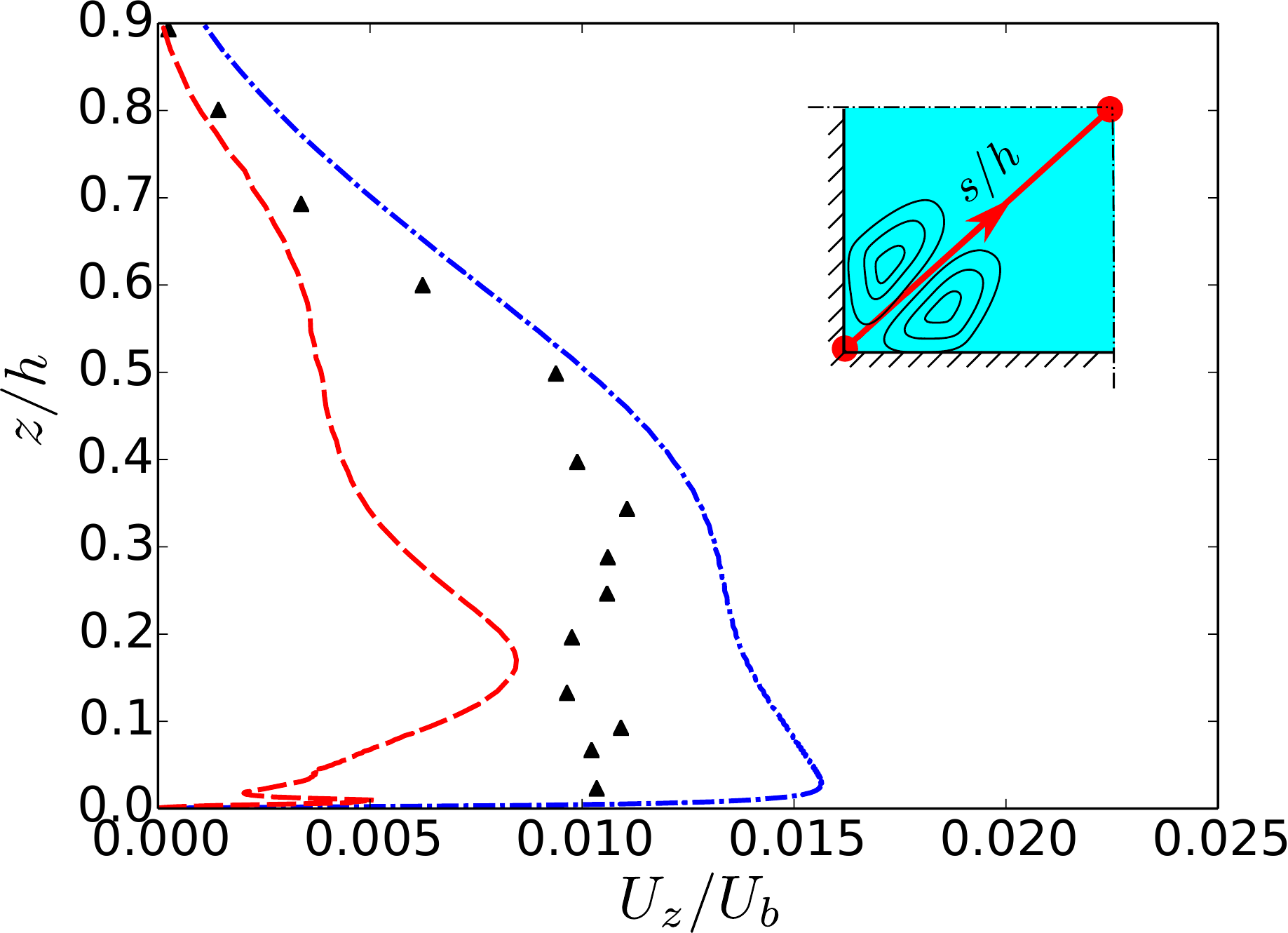}} \hspace{0.2em}
\caption{The comparison of predicted secondary velocity $U_z$ at Reynolds number $Re=1.25\times10^5$ with experimental data~\cite{gessner81numerical} (denoted as $\blacktriangle$). Comparisons are shown (a) along the vertical axis of symmetry and (b) along the diagonal of the duct. The training flow is at Reynolds number $Re=2200$.}
\label{fig:U_duct_25E4}
\end{figure}

To further demonstrate the general applicability of the proposed framework, we employ the same training flow and investigate another test flow, i.e., the flow in a square duct at a much higher Reynolds number $Re=1.25 \times 10^5$. We evaluate the prediction performance of the proposed framework by using the experimental data of this test flow along the vertical axis of symmetry and along the diagonal of the square duct~\cite{gessner81numerical}. The inverse flow near the bottom of the axis of symmetry is not captured by the baseline RANS simulation in Fig.~\ref{fig:U_duct_25E4}a. This inverse flow in experimental data indicates that there is a small vortex with an opposite rotation direction around this region, in addition to the main vortex of the secondary flow. The failure of predicting the inverse flow in Fig.~\ref{fig:U_duct_25E4}a means that this small vortex is completely missing in the baseline RANS simulation results. In contrast, this missing flow characteristic is successfully captured by the machine-learning prediction. It should be noted that this flow characteristic is not observed in the training flow at a much lower Reynolds number $Re=2200$. One reason for the successful prediction of the small inverse flow is the formulation of our data-driven augmentation framework, i.e., it is the Reynolds stress discrepancies but not the whole Reynolds stress that is predicted by the machine learning framework. Specifically, Wu et al.~\cite{wu16bayesian} calibrated the Reynolds stress discrepancies at a lower Reynold number $Re\approx5000$ and applied the calibrated discrepancies to correct the RANS simulation at a higher Reynolds number $Re=125000$. They reported that the trend of inverse flows could be re-produced even though the Reynolds stress discrepancies are calibrated at a much lower Reynolds number. Therefore, the trend of small inverse flows can be predicted even if the machine learning model is overfitted for the training database at a lower Reynolds number. However, it should be noted that the inverse flow was noticeably underestimated in the work by Wu et al.~\cite{wu16bayesian}, and the machine learning prediction in this work achieves better agreement with the experimental data. Therefore, the successful prediction of the inverse flow in Fig.~\ref{fig:U_duct_25E4}a is a strong evidence that the machine-learning-assisted turbulence modeling indeed has the potential of revealing the physics within the data, rather than simply interpolating with the available data.

As shown in Fig.~\ref{fig:U_duct_25E4}b, the secondary flow $U_z$ is over-predicted by baseline RANS simulation along the diagonal of the duct. On the contrary, the mean velocity $U_z$ obtained by the proposed framework underestimates the magnitude of the secondary flow in Fig.~\ref{fig:U_duct_25E4}b. Although there is no significant improvement of results along the diagonal, it should be noted that the machine learning prediction in Fig.~\ref{fig:U_duct_25E4}b indeed corrects the baseline RANS simulated results towards the right direction. 

\subsubsection{Flow over periodic hills}
\label{sec:pehill}
\begin{figure}[htbp]
\centering
\includegraphics[width=0.3\textwidth]{Tau-legend.pdf} \\
\subfloat[$\bm{\tau}_{xy}$]{\includegraphics[width=0.485\textwidth]{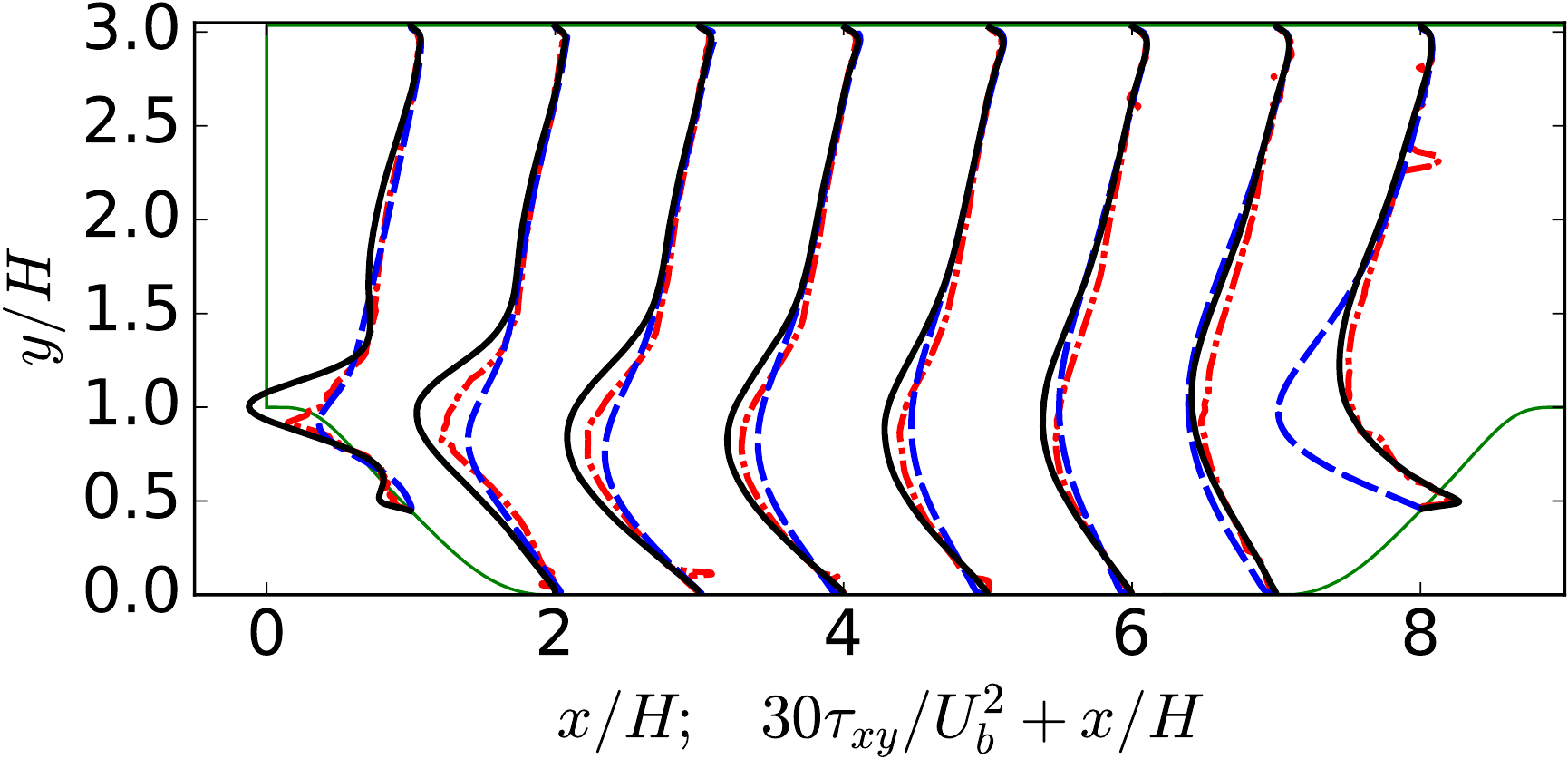}} \hspace{0.5em}
\subfloat[Turbulent Kinetic Energy]{\includegraphics[width=0.485\textwidth]{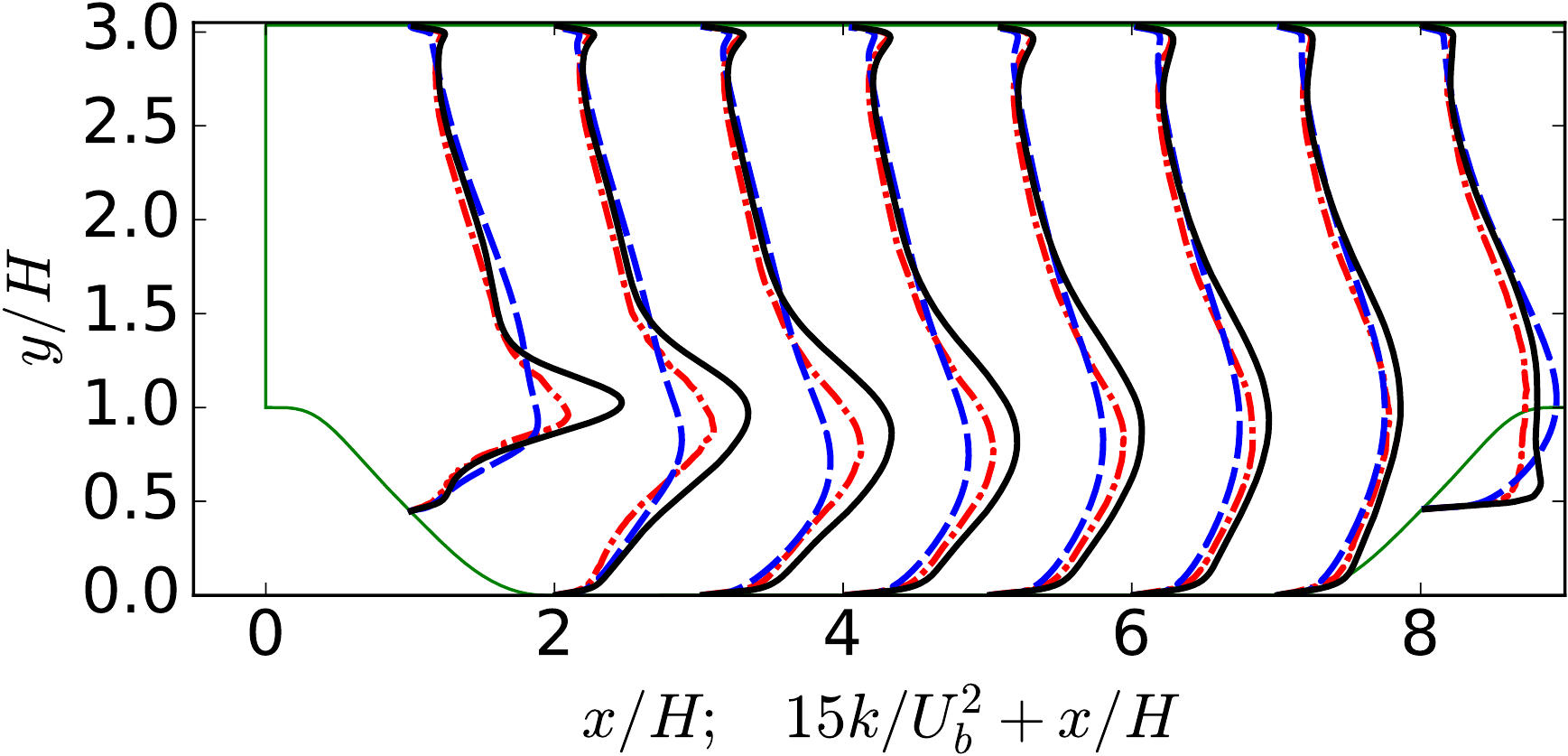}}
\caption{The prediction of (a) shear stress component $\bm{\tau}_{xy}$ of the \textbf{Reynolds stress} tensor and (b) the turbulent kinetic energy (TKE)  at $x/H=1, 2, \ldots ,8$. The test flow is the flow over periodic hill at $Re=5600$. The training flow is at the same Reynolds number but has a steeper hill profile as shown in Fig.~\ref{fig:domain_pehill}.}
\label{fig:Tau_xy_pehill}
\end{figure}

In the cases investigated above, the training flow and the test flow share the same geometry configuration and only differ in Reynolds numbers. In order to demonstrate the capability of the proposed framework for the flows with different geometries, we further study the flows over periodic hills with different shape of hill profiles. The training flow is the flow over periodic hill at $Re=5600$~\cite{breuer09flow}, and the test flow is also at $Re=5600$ but has a steeper hill profile as described in Sec.~\ref{sec:case-setup}. It can be seen in Fig.~\ref{fig:Tau_xy_pehill}a that the RANS simulation under-predicts the shear stress $\tau_{xy}$ at downstream of the hill crest. Such under-prediction of shear stress is mainly due to the under-prediction of TKE of RANS simulation as shown in Fig.~\ref{fig:Tau_xy_pehill}b. Compared with the results of RANS simulation, the machine learning prediction shows a better agreement with DNS data for both the shear stress and the TKE. Although such improvement is limited at $x/H=1$, it still better predicts the magnitude of Reynolds stress than RANS simulation as shown in Fig.~\ref{fig:Tau_xy_pehill}. The improvement of the machine learning prediction at further downstream becomes more prominent.

\begin{figure}[htbp]
\centering
\hspace{2em}\includegraphics[width=0.3\textwidth]{Tau-legend.pdf} \\
\includegraphics[width=0.7\textwidth]{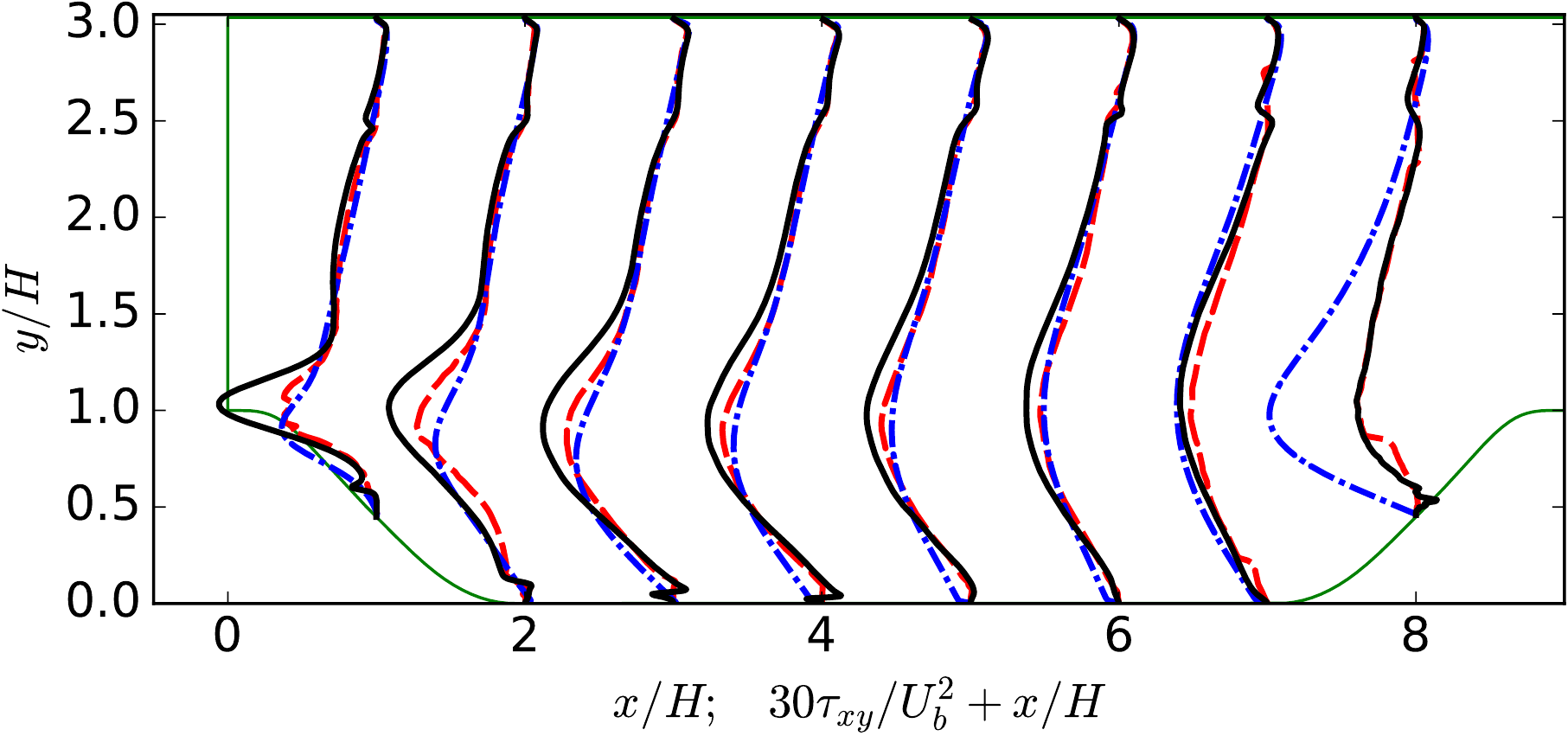}
\caption{The prediction of shear stress component $\bm{\tau}_{xy}$ of \textbf{linear part of Reynolds stress} at $x/H=1, 2,...,8$. The test flow is the flow over periodic hill at $Re=5600$. The training flow is at the same Reynolds number but has a steeper hill profile as shown in Fig.~\ref{fig:domain_pehill}.}
\label{fig:TauLS_xy_pehill}
\end{figure}

The linear part of Reynolds stress $\bm{\tau}^{L}$ is also predicted. It can be seen in Fig.~\ref{fig:TauLS_xy_pehill} that the linear part of Reynolds stress is similar to the DNS Reynolds stress in Fig.~\ref{fig:Tau_xy_pehill} except for several regions, e.g., the region around the bottom wall and the region within upper channel around $y/H=2.5$. Therefore, the nonlinear part of the DNS Reynolds stress is expected to be less dominant in most regions for the flow over periodic hills. The prediction of the proposed framework shows a good agreement with the linear part of Reynolds stress obtained from DNS data. Similar to the prediction of Reynolds stress, the machine learning prediction of the linear part of Reynolds stress shows less noticeable improvement at $x/H=1$.

\begin{figure}[htbp]
\centering
\hspace{2em}\includegraphics[width=0.2\textwidth]{nut-legend.pdf} \\
\includegraphics[width=0.8\textwidth]{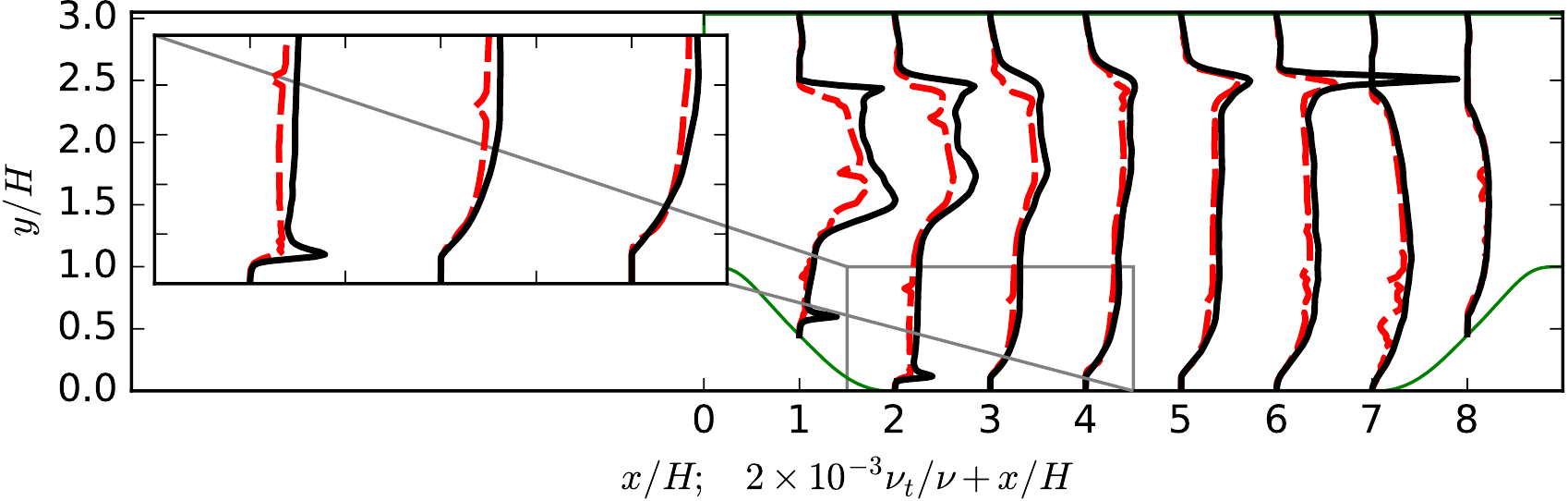}
\caption{The machine-learning-predicted optimal eddy viscosity at $x/H=1, 2,...,8$. The test flow is the flow over periodic hill at $Re=5600$. The training flow is at the same Reynolds number but has a steeper hill profile as shown in Fig.~\ref{fig:domain_pehill}.}
\label{fig:nutLS_pehill}
\end{figure}

In addition to the improvement shown in the prediction of DNS Reynolds stress and its linear part, the prediction of optimal eddy viscosity also demonstrates improvement as shown in Fig.~\ref{fig:nutLS_pehill}. It can be seen in Fig.~\ref{fig:nutLS_pehill} that the machine-learning-predicted eddy viscosity has a good agreement with the DNS eddy viscosity, except for a few regions where the DNS eddy viscosity changes rapidly. Such deterioration in the performance of machine learning prediction is expected, since the functions with such behavior pose more difficulties in machine learning. However, the peak value of eddy viscosity at these regions is usually not important for solving for mean velocity. It is because the strain rate tensor is close to zero in these regions, corresponding to the peak value of eddy viscosity.

\begin{figure}[htbp]
\centering
\includegraphics[width=0.65\textwidth]{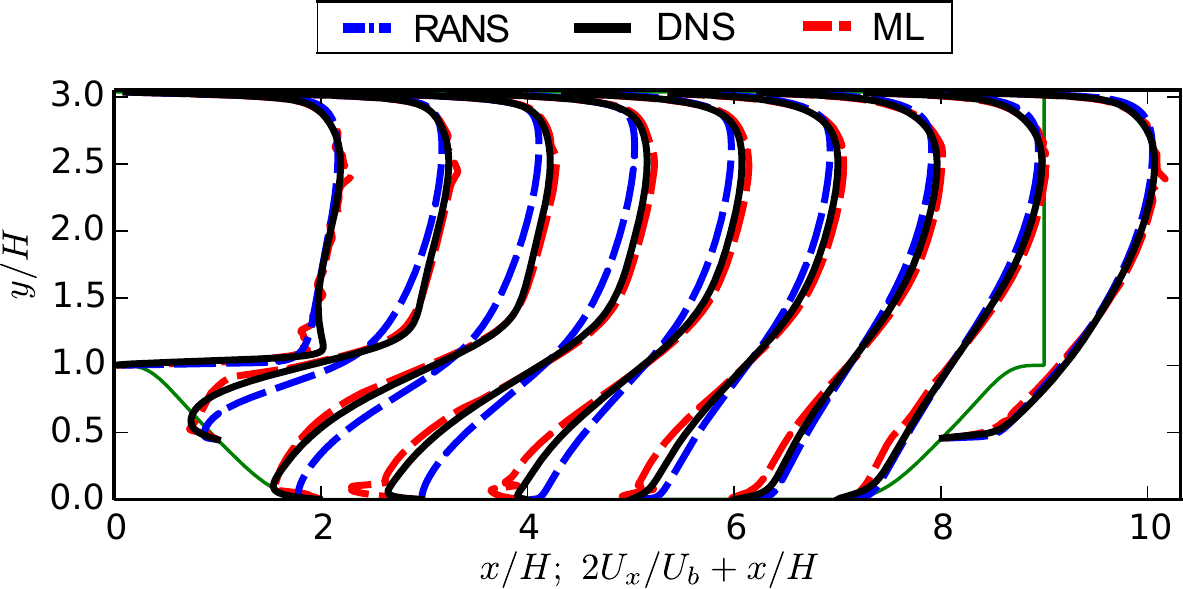}
\caption{The stream-wise velocity by RSM with implicit treatment at $x/H=1, 2,...,8$. The test flow is the flow over periodic hill at $Re=5600$. The training flow is at the same Reynolds number but has a steeper hill profile as shown in Fig.~\ref{fig:domain_pehill}.}
\label{fig:Ux_pehill}
\end{figure}

The comparison of mean velocity field in Fig.~\ref{fig:Ux_pehill} shows that the mean velocity obtained by the proposed framework has a better agreement with the DNS data. Specifically, the reverse flow extends to $x/H=4$ in DNS data, denoting the size of the separation bubble downstream of the hill crest. The RANS simulation results indicate that the reverse flow ends approximately around $x/H=3$, which significantly underestimate the size of the separation bubble. The magnitude of velocity at upper channel region is also under-predicted by the RANS simulation results from $x/H=1$ to $x/H=5$. Compared with the RANS simulation results, the machine learning prediction provides more accurate reverse flow, especially from $x/H=1$ to $x/H=3$. Although over-prediction of reverse flow can be observed in the prediction of the proposed framework from $x/H=3$ to $x/H=4$, the separation region is still better predicted than RANS simulation results. In addition, the proposed framework achieves a much better prediction of mean velocity than the RANS simulation results at the upper channel region.

\section{Discussion: Potentials and Limitations of Data-Driven Turbulence Models}
\label{sec:discussion}
The present work addresses the ill-conditioning issue in a class of data-driven turbulence models that aim to model the Reynolds stresses with machine learning.  Therefore, it is helpful to provide a broad yet brief view of data-driven turbulence modeling, which has emerged as a promising yet controversial subject in the past few years. A more comprehensive state-of-the-art overview  will be presented in a forthcoming review article~\cite{ARFM}. Summarized briefly, three distinctly different approaches to data-driven turbulence modeling have been pursued by different groups:
\begin{enumerate}[(i)]%
\item Weatheritt and Sandsberg~\cite{weatheritt16novel,weatheritt17development} used Gene Expression Programming to develop algebraic Reynolds stress models based on symbolic regression.
\item Duraisamy and co-workers~\cite{singh17machine,singh17data} used machine learning to predict discrepancies in the source terms in the turbulence transport equations for existing models (e.g., S--A model, $k$--$\omega$ model, or Reynolds stress models).
\item Wang et al.~\cite{wang17physics-informed} and Ling et al.~\cite{ling16reynolds} used machine learning to directly predict the Reynolds stresses or their discrepancies compared to the truth.
\end{enumerate}
A widely accepted yardstick used in the turbulence modeling community to assess turbulence models is that an ideal model should
\begin{itemize}
\item demonstrate robust predictive capabilities in a wide range of flows without flow-specific tuning
\item be interpretable, ideally in explicit analytical forms
\end{itemize}
which are referred to as \emph{universality} and \emph{interpretability} requirements, respectively, hereafter. The three approaches above are discussed in light of each of these requirements. 

\subsection{Universality of turbulence models}
In terms of the universality requirement, existing traditional turbulence models are still far from satisfactory. To our knowledge, most existing data-driven turbulence models, including the one presented in our work, are still in their infancy and have shown only limited predictive capabilities, typically in flows that are close to the training flows. That is, the training and predictions flows belong to the same class of flows (e.g., massively separated flows) but with variations of flow configurations such as geometry or Reynolds number. 
However, their predictive capabilities may improve as more experiences are accumulated and the methodologies are refined later on. The machine learning inputs and outputs in this work are all rotational invariants and the Galilean invariants. Therefore, the trained model is objective under any rotation transformation and Galilean transformation. Although several inputs and outputs are not reflection invariants, the associated extrapolation capability can be achieved by reflecting the coordinates system to augment the training data. However, the extrapolation capability still depends on the diversity of the training data. By using the training database of flows with a specific characteristic, e.g., recirculation or stress-induced secondary flow, we have demonstrated that an unknown flow with similar characteristics can be predicted. We envision that the extrapolation capability can be extended to more complex flows with more diversified and systematically generated high fidelity simulation database. In addition, the extrapolation capability for more complex flows can also be achieved by automatically classifying the flow field into different regions and making prediction for each region accordingly. On the other hand, if we take a less ambitious perspective and consider the data-driven models (particularly the second and third approaches) as augmentation of traditional models through a data-driven, flow-specific correction, then their universality become less critical. For example, in a practical applications, a data-driven model could turn itself off when the part of the flow is not present in the training dataset and is too ``far'' from the flows in the data. That is, the data-driven model can revert back to the traditional, baseline turbulence model that is used. To this end, Wu et al.~\cite{wu17priori} has demonstrated a fully automatic, statistically rigorous way of measuring the ``distance'' between two flows (or any parts thereof) based on the mean flow fields obtained from RANS simulations. With such a distance metric, it is possible to ensure that a machine-learning-augmented model is at least better than or the same as the corresponding baseline model it utilizes.

\subsection{Interpretability of turbulence models}
Regarding the interpretability requirement, it may appear that the symbolic regression approach specifically aims to find Reynolds stress models that are in analytical forms, while the second and third approaches based on machine learning (neural networks and random forests) do not produce models in analytical forms and are not amenable for interpretation. However, the delineation between these approaches may not be as clear as it first appears. First, although the analytical Reynolds stress model learned from one class of flows~\cite{weatheritt16novel} may rather simple and indeed has similar complexity to existing advanced turbulence models, this may not necessarily be true for other flows. For example, a model learned from a diverse datasets from different flows may be an analytical expression with the number of terms too large for human comprehension, which may still be difficult to interpret despite its analytical nature. Second, if one is willing to compromise the predictive performance, a neural network based model (used in the second and third approaches) may be regularized (e.g., by using Ridge and LASSO~\cite{james13introduction}) to yield simple architecture and thus becomes more interpretable. Finally, both neural networks and random forest can provide the importance of input mean flow features~(see, e.g. the discussions in \cite{wang17physics-informed}), which may help traditional model developers incorporate additional variables to existing models. Efforts along such a direction are underway~\cite{durbin}.
Therefore, the fundamental differences among the three approaches above lie in the classical trade-off between predicative capability and interpretability in models~\cite{james13introduction}, and not in the specific forms of the chosen  model (e.g., symbolic regression v.s. neural networks or random forests).

Ultimately, both universality and interpretability requirements are intimately related to the fundamental question in turbulence modeling: does there exist a universal turbulent constitutive relation?  Generations of researchers have labored for many decades on dozens of turbulence models, yet none of them achieved predictive generality, which seems to indicate that the answer is ``no''. If so, then flow-specific tuning and fudge factors would be inevitable if good predictive performances are desirable. The machine learning based turbulence models can be considered automatic, flow-specific tuning schemes based on the flow regime to be predicted and the flow regimes that present in the training database.

\section{Conclusion}
\label{sec:conclusion}
While earlier works demonstrated the capabilities of machine learning in predicting improved Reynolds stresses, obtaining improved mean velocity field remains a challenge of machine-learning-assisted turbulence modeling. The main reason is the sensitivity of mean velocity with regard to the errors in the prediction of Reynolds stress. In this work we propose a physics-based implicit treatment to model Reynolds stress by using machine learning techniques. Specifically, the optimal eddy viscosity and the nonlinear part of Reynolds stress are both predicted. In the propagation test, the DNS Reynolds stress is used in solving for mean velocity to illustrate the ideal scenario of machine-learning-assisted turbulence modeling approaches. The propagation test with DNS Reynolds stress shows that satisfactory mean velocity can be achieved by the Reynold stress models with implicit treatment. In the \textit{a posteriori} test, three training-prediction cases are investigated to demonstrate the predictive capability of the proposed framework. In the first and second cases, a machine learning model is trained on the flow in a square duct at Reynolds number $Re=2200$, and the flows in a square duct at Reynolds numbers $Re=3500$ and $Re=1.25 \times 10^5$ are predicted. In the third case, machine learning model is trained on the flow over periodic hills at Reynolds number $Re=5600$, and the flow with a steeper hill profile is predicted. The satisfactory prediction performance of mean velocity field demonstrates the predictive capability of the proposed machine-learning-assisted framework. Specifically, machine-learning-trained model successfully predicts the mean flow pattern in the second case that is not even shown in the training flow. It provides a strong evidence that machine-learning-assisted turbulence modeling can reveal flow physics from the existing data, instead of merely fitting on the existing data. With the capability in predicting the mean velocity field, the proposed physics-based implicit treatment leads to a practical machine-learning-assisted turbulence modeling framework in real applications where mean velocity field and other  quantities of interests need to be predicted.

\section{Acknowledgment}
The authors would like to thank the two reviewers for their constructive and valuable comments, which helped improving the quality and clarity of this manuscript. In particular, one of the reviewers pointed out the lack of Galilean invariance in two of the normalization constants in our manuscript, which we fixed during the revision. We gratefully acknowledge the reviewer's constructive comments.

\appendix
\section{Data-Driven Reynolds Stress Model: Detailed Algorithms}
\label{sec:workflow}
The detailed work flow of constructing a data-driven Reynolds stress model with implicit treatment and use it for solving the RANS  equations are presented as follows. First, 
we write Eq.~\ref{eq:b} in terms of the Reynolds stress tensor rather than anisotropy tensor:
\begin{equation}
\label{eq:b2}
\mathbf{b}=\nu_t^{L} \mathbf{S} + (\mathbf{b} - \nu_t^{L} \mathbf{S})=\nu_t^{L} \mathbf{S} + (\bm{\tau} - (\nu_t^{L} \mathbf{S}+\textrm{tr}(\bm{\tau})))=\nu_t^{L} \mathbf{S} + (\bm{\tau} - \bm{\tau}^{L})
\end{equation}
where $\textrm{tr}(\bm{\tau})$ represents the trace of the Reynolds stress tensor, and $\bm{\tau}^{L}$ denotes the linear part of Reynolds stress tensor.

The procedure of the proposed framework are as follows:
 \begin{enumerate}[(i)]
 	\item Perform baseline RANS simulations on both the training flows and the test flow to obtain mean flow features $\mathbf{q}$ and Reynolds stress tensor $\bm{\tau}^{\textrm{RANS}}$.
	\item Train regression functions for Reynolds stress discrepancies $\Delta \bm{\tau}$ and $\Delta \bm{\tau}^{L}$, and use the trained regression functions to predict the test flow.
	\begin{enumerate}
 	\item Compute the discrepancies fields $\Delta \bm{\tau}=\bm{\tau}^{\textrm{DNS}}-\bm{\tau}^{\textrm{RANS}}$ and $\Delta \bm{\tau}^{L}=\bm{\tau}^{L}-\bm{\tau}^{\textrm{RANS}}$ for the training flows based on the DNS data, and construct regression functions $ f_1: \mathbf{q} \mapsto \Delta \bs\tau$ and $ f_2: \mathbf{q} \mapsto \Delta \bm{\tau}^{L}$ by using PIML framework~\cite{wang17physics-informed}.
 	\item Use trained regression functions $f_1$ and $f_2$ to predict the discrepancies fields $\Delta \bm{\tau}$ and $\Delta \bm{\tau}^{L}$ for the test flow, and compute the corresponding Reynolds stress fields $\bm{\tau}$ and $\bm{\tau}^{L}$ by adding the predicted discrepancies fields to the RANS simulated Reynolds stress $\bm{\tau}^{\textrm{RANS}}$.
	\end{enumerate}
        \item Train regression functions for eddy viscosity $\nu_t^{L}$, and use the trained regression functions to predict for the test flow.
        \begin{enumerate}
	\item Compute the least squares eddy viscosity $\nu_t^{L}$ for the training flows based on the DNS data, and construct regression function $ f_3: \mathbf{q} \mapsto \nu_t^{L}$ by using machine learning.
	\item Use trained regression functions $f_3$ to predict the least squares eddy viscosity $\nu_t^{L}$ for the test flow. 
	\end{enumerate}
	\item Substitute the predicted $\nu_t^{L}$, $\bm{\tau}$ and $\bm{\tau}^{L}$ into RANS equations for the anisotropy stress tensor as shown in Eq.~\ref{eq:b2}, and solve for the corresponding mean velocity field.
 \end{enumerate}
 
In this workflow, the linear part and nonlinear part of Reynolds stress are predicted separately. For the linear part, we only predict the optimal eddy viscosity $\nu_t^L$ and thus treat the linear part implicitly. To obtain the nonlinear part, we predict both the Reynolds stress itself and the linear part of Reynolds stress, and we calculate the nonlinear part of Reynolds stress by subtracting the linear part of Reynolds stress from the Reynolds stress itself.
 
The solving for mean velocity are performed in a finite-volume CFD platform OpenFOAM, using a modified flow solver that allows the implicit treatments of Reynolds stress predicted by the machine learning model. Specifically, the modified flow solver is based on a
built-in steady-state incompressible flow solver \texttt{simpleFoam}~\citep{weller98tensorial}, in
which the SIMPLE algorithm~\cite{patankar80numerical} is used. Unlike the standard \texttt{simpleFoam} solver, the modified flow solver uses the machine learning predicted $\nu_t^{L}$, $\bm{\tau}$ and $\bm{\tau}^{L}$ to represent the modeled Reynolds stress $\bm{\tau}^m$, i.e. $\bm{\tau}^m=\nu_t^{L} \mathbf{S} + (\bm{\tau} - \bm{\tau}^{L})+\textrm{tr}(\bm{\tau})$. The strain rate tensor $\mathbf{S}$ is treated implicitly in the modified flow solver. For numerical discretizations of the RANS equations, the second-order central difference scheme is chosen for all terms except for the convection term, which is discretized with a second-order upwind scheme.

For the boundary conditions of the machine learning predicted $\nu_t^{L}$, $\bm{\tau}$ and $\bm{\tau}^{L}$ in this work, the fixed zero values are applied at the walls and periodic boundary conditions are applied in the streamwise ($x$) direction. The boundary conditions of the mean velocity and the pressure are treated the same as the standard RANS simulations, i.e., periodic boundary conditions are applied in the streamwise ($x$) direction, 
and non-slip boundary conditions are applied at the walls.

\section{Integrity Basis of Mean Flow Features}
\label{sec:list-of-basis}
\begin{table}[htbp]  
	\centering
	\caption{Minimal integrity basis for symmetric tensor $\widehat{\mathbf{S}}$ and antisymmetric tensors $\widehat{\bs{\Omega}}$, $\widehat{\mathbf{A}}_{p}$, and $\widehat{\mathbf{A}}_{k}$. In the implementation, $\widehat{\mathbf{S}}$ is the strain-rate tensor, $\widehat{\bs{\Omega}}$ is the rotation-rate tensor; $\widehat{\mathbf{A}}_{p}$ and $\widehat{\mathbf{A}}_{k}$ are the antisymmetric tensors associated with 
	pressure gradient $\widehat{\nabla p}$ and the gradient of turbulence kinetic energy $\widehat{\nabla k}$;
	$n_S$ and $n_A$ denote the numbers of symmetric and antisymmetric raw tensors for the integrity basis. 
	}
	\label{tab:basis}
	\begin{tabular}{c|C{2.5cm}|C{9.5cm}}	
		\hline
		$(n_S, n_A)$ &  feature index &  invariant bases$^{(\mathrm{a})}$\\
		\hline
		(1, 0) & 1--2 & $\widehat{\mathbf{S}}^2$, $\widehat{\mathbf{S}}^3$ \\
		\hline
		(0, 1)& 3--5 & $\widehat{\bs{\Omega}}^2$,  $\widehat{\mathbf{A}}_{p}^2$,  $\widehat{\mathbf{A}}_{k}^2$ \\
		\hline
		\multirow{3}{*}{(1, 1)} & \multirow{3}{*}{6--14} 
		& $\widehat{\bs{\Omega}}^2 \widehat{\mathbf{S}}$, $\widehat{\bs{\Omega}}^2 \widehat{\mathbf{S}}^2$, $\widehat{\bs{\Omega}}^2 \widehat{\mathbf{S}} \widehat{\bs{\Omega}} \widehat{\mathbf{S}}^2$;\\
		&& $\widehat{\mathbf{A}}_{p}^2 \widehat{\mathbf{S}}$, $\widehat{\mathbf{A}}_{p}^2 \widehat{\mathbf{S}}^2$, $\widehat{\mathbf{A}}_{p}^2 \widehat{\mathbf{S}} \widehat{\mathbf{A}}_{p} \widehat{\mathbf{S}}^2$;\\
		&& $\widehat{\mathbf{A}}_{k}^2 \widehat{\mathbf{S}}$, $\widehat{\mathbf{A}}_{k}^2 \widehat{\mathbf{S}}^2$, $\widehat{\mathbf{A}}_{k}^2 \widehat{\mathbf{S}} \widehat{\mathbf{A}}_{k} \widehat{\mathbf{S}}^2$; \\
		\hline
		(0, 2)& 15--17 & $\widehat{\bs{\Omega}} \widehat{\mathbf{A}}_{p}$, $\widehat{\mathbf{A}}_{p} \widehat{\mathbf{A}}_{k}$, $\widehat{\bs{\Omega}} \widehat{\mathbf{A}}_{k}$ \\
		\hline
		\multirow{3}{*}{(1, 2)} & \multirow{3}{*}{18--41} & 
		$\widehat{\bs{\Omega}} \widehat{\mathbf{A}}_{p} \widehat{\mathbf{S}}$, $\widehat{\bs{\Omega}} \widehat{\mathbf{A}}_{p} \widehat{\mathbf{S}}^2$, $\widehat{\bs{\Omega}}^2 \widehat{\mathbf{A}}_{p} \widehat{\mathbf{S}}$*, $\widehat{\bs{\Omega}}^2 \widehat{\mathbf{A}}_{p} \widehat{\mathbf{S}}^2$*, $\widehat{\bs{\Omega}}^2 \widehat{\mathbf{S}} \widehat{\mathbf{A}}_{p} \widehat{\mathbf{S}}^2$*;\\
		
		&&$\widehat{\bs{\Omega}} \widehat{\mathbf{A}}_{k} \widehat{\mathbf{S}}$,  $\widehat{\bs{\Omega}} \widehat{\mathbf{A}}_{k} \widehat{\mathbf{S}}^2$,
		$\widehat{\bs{\Omega}}^2 \widehat{\mathbf{A}}_{k} \widehat{\mathbf{S}}$*,  $\widehat{\bs{\Omega}}^2 \widehat{\mathbf{A}}_{k} \widehat{\mathbf{S}}^2$*, $\widehat{\bs{\Omega}}^2 \widehat{\mathbf{S}} \widehat{\mathbf{A}}_{k} \widehat{\mathbf{S}}^2$*;\\
		
		&&$\widehat{\mathbf{A}}_{p} \widehat{\mathbf{A}}_{k} \widehat{\mathbf{S}}$, $\widehat{\mathbf{A}}_{p} \widehat{\mathbf{A}}_{k} \widehat{\mathbf{S}}^2$,
		$\widehat{\mathbf{A}}_{p}^2 \widehat{\mathbf{A}}_{k} \widehat{\mathbf{S}}$*, $\widehat{\mathbf{A}}_{p}^2 \widehat{\mathbf{A}}_{k} \widehat{\mathbf{S}}^2$*, $\widehat{\mathbf{A}}_{p}^2 \widehat{\mathbf{S}} \widehat{\mathbf{A}}_{k} \widehat{\mathbf{S}}^2$*;\\
		\hline
		(0, 3) & 42 & $\widehat{\bs{\Omega}} \widehat{\mathbf{A}}_{p} \widehat{\mathbf{A}}_{k}$ \\
		\hline
		(1, 3) & 43--47 &   $\widehat{\bs{\Omega}} \widehat{\mathbf{A}}_{p} \widehat{\mathbf{A}}_{k} \widehat{\mathbf{S}}$,  $\widehat{\bs{\Omega}} \widehat{\mathbf{A}}_{k} \widehat{\mathbf{A}}_{p} \widehat{\mathbf{S}}$,  $\widehat{\bs{\Omega}} \widehat{\mathbf{A}}_{p} \widehat{\mathbf{A}}_{k} \widehat{\mathbf{S}}^2$,
		$\widehat{\bs{\Omega}} \widehat{\mathbf{A}}_{k} \widehat{\mathbf{A}}_{p} \widehat{\mathbf{S}}^2$,  $\widehat{\bs{\Omega}} \widehat{\mathbf{A}}_{p} \widehat{\mathbf{S}} A_3 \mathbf{S}^2$ \\
		\hline						 								
	\end{tabular}
	\flushleft
	{\small
		Note: (a) The invariance basis is the trace of each tensor listed below. \\
        (b) The asterisk ($*$) on a term indicates that all terms formed by cyclic
 permutation of anti-symmetric tensor labels (e.g., $\widehat{\bs{\Omega}}^2 \widehat{\mathbf{A}}_{p} \widehat{\mathbf{S}}$* is short for $\widehat{\bs{\Omega}}^2 \widehat{\mathbf{A}}_{p} \widehat{\mathbf{S}}$ and $ \widehat{\mathbf{A}}_{p}^2 \widehat{\bs{\Omega}} \widehat{\mathbf{S}}$) are also included.

        }	
\end{table}
The minimal integrity bases for rotational invariance with given input 
set \[\widehat{\mathcal{Q}} = \{\widehat{\mathbf{S}}, \widehat{\bm{\Omega}}, \widehat{\nabla p}, \widehat{\nabla k}\}\] of tensors is shown in Table~\ref{tab:basis}. Spencer~\cite{spencer62isotropic,johnson16handbook} provided a systematic procedure of generating minimal invariant bases from a set of symmetric and anti-symmetric tensors. In order to utilize this procedure, we need to first transform the vectors $\widehat{\nabla p}$ and $\widehat{\nabla k}$ to the corresponding antisymmetric tensors by using the following mapping~\cite{johnson16handbook}:
 \begin{subequations}
 	\label{eq:vector2anti}
 	\begin{align}
 	\widehat{\mathbf{A}}_p & =  -\mathbf{I} \times \widehat{\nabla p}\\
 	\widehat{\mathbf{A}}_k & =  -\mathbf{I} \times \widehat{\nabla k}
 	\end{align}  	 
 \end{subequations}
 where $\mathbf{I}$ is the second order identity tensor, and the symbol $\times$ denotes tensor cross
 product. 

\section{Galilean Invariance}
\label{app:gali-inv}
In Section~\ref{sec:meth-invariance}, we stated that the all feature variables in Table~\ref{tab:featureRaw} and~\ref{tab:feature} and their corresponding normalization factors are Galilean invariant. This is evidence from the fact that most of the variables and normalization factors contains only terms associated with the velocity gradient $\nabla \mathbf{U}$ (e.g., $\mathbf{S}$ and $\mathbf{\Omega}$), pressure gradient $\nabla p$, and TKE gradient $\nabla k$. However, the Galilean invariance of the normalization facor $\rho |Du/Dt|$ is not evident.  In this appendix we show that the material derivative of velocity is Galilean invariant.

The mean velocity at location $\bm{x}$ and time $t$ observed in a stationary reference frame is denoted as $\mathbf{U}(\bm{x},t)$. The mean velocity observed in reference frame moving with constant velocity $\mathbf{C}$ can be written as $\mathbf{U}^{*} (\bm{x}^*,t) = \mathbf{U}(\bm{x}-\mathbf{C}t, t) + \mathbf{C}$, where $\bm{x}^*=\bm{x} - \mathbf{C}t$ represents the spatial location observed in the moving reference frame. The material derivative of the velocity $U^*_j$ in the moving reference frame is derived as follows:
\begin{equation}
\label{eq:dudt1}
\frac{\partial U_i^*}{\partial t}=\frac{\partial U_i}{\partial t}-C_j\frac{\partial U_i^*}{\partial x_j^*}
\quad \text{and} \quad
U_j^*\frac{\partial U_i^*}{\partial x_j^*}=(U_j+C_j)\frac{\partial U_i^*}{\partial x_j^*}
\end{equation}
Combining the two terms in Eq.~\ref{eq:dudt1} and  utilizing the fact that $\partial U_i^*/\partial x_j^*=\partial U_i/\partial x_j$ (because the reference frame velocity is constant) yields:
\begin{equation}
\frac{\partial U_i^*}{\partial t}+U_j^*\frac{\partial U_i^*}{\partial x_j^*}=\frac{\partial U_i}{\partial t}+U_j\frac{\partial U_i}{\partial x_j} ,
\end{equation}
That is,
\begin{equation}
\frac{D U_i^*}{D t}  = \frac{D U_i}{Dt} 
\end{equation}
which demonstrates that the material derivative of velocity $\mathbf{U}$ is invariant under Galilean transformation. The merit of ensuring Galilean invariance is that the trained machine learning model $g(\mathbf{U})$ is valid in all inertial frames, i.e., $g(\mathbf{U})=g(\mathbf{U}+\mathbf{C})$, enhancing the generality of the trained model.

It is noted that several input features in the works of Wang et al.~\cite{wang17physics-informed} and Ling and Templeton~\cite{ling15evaluation} are not Galilean invariant. For example, the raw features of pressure gradient along streamline $U_i \partial P /\partial x_i$ and the ratio of convection to production of turbulent kinetic energy $U_i dk/dx_i$, and the normalization factor $U_iU_i$ for turbulence intensity and the normalization factor $\rho \partial U_i^2/\partial x_i$ for the ratio of pressure normal stresses to shear stresses.  Therefore, the machine learning framework in this work is expected to achieve better extrapolation capability with the Galilean invariance for all input and output quantities.

\clearpage
\bibliographystyle{elsarticle-num}

\end{document}